\documentclass[letterpaper]{article}
\usepackage{amsfonts,amssymb,amsmath,bm,bbm,euscript,mathrsfs,calligra}
\usepackage[usenames]{color}
\usepackage{color,soul}
\usepackage[utf8]{inputenc}
\usepackage[T1]{fontenc}
\usepackage{textcomp}
\usepackage{graphicx}
\usepackage[tight,sf]{subfigure}
\usepackage{listings}
\usepackage{hyperref}
\usepackage{bookmark}
\usepackage{authblk}

\DeclareMathAlphabet{\mathcalligra}{T1}{calligra}{m}{n}
\DeclareFontShape{T1}{calligra}{m}{n}{<->s*[2.2]callig15}{}

\begin{document}

\sf
\begin{center}
   \vskip 2em
  {\huge \sf  Conformal Mechanics of Planar Curves}

\vskip 3em
{\large \sf  Jemal Guven and Gregorio Manrique}

\vskip 2em

\em{Instituto de Ciencias Nucleares\\
Universidad Nacional Aut\'onoma de M\'exico\\
Apdo. Postal 70-543, 04510 Ciudad de M\'exico,  MEXICO}
\end{center}

 \vskip 1em

\begin{abstract}
\sf 
Self-similar curves arise naturally as the tension-free equilibrium states of 
conformally invariant bending energies. The simplest example is the M\"obius invariant conformal arc-length on planar curves, dependent on the Frenet curvature $\kappa$ through its first derivative with respect to 
arc-length. There are four conserved currents associated with this invariance: 
the  tension and torque associated with Euclidean invariance,  as well as scalar and vector currents reflecting invariance under scaling and special conformal transformations respectively.  
If the tension vanishes, all equilibrium states are self-similar:
in the case of conformal arc-length, these are logarithmic spirals with no internal structure.  More generally, the tension-free states are logarithmic spirals decorated with  a repeating self-similar internal structure.  Here it will be shown how the conservation laws can be used to construct these curves, while also endowing their geometry with a mechanical interpretation. The scaling current and the  
torque together provide a scale-invariant ode for the dimensionless variable $\kappa'/\kappa^2$, which captures the internal structure of the spiral.  For conformal arc-length it is constant.  In tension-free states, the special conformal current vanishes. Its projections along orthogonal directions determine directly the distance from the spiral apex locally in terms of the curvature. 
The  
quadratic Casimir invariant of the M\"obius group can be cast in terms of the four currents, none of which itself is invariant.  For conformal arc-length, this is identified as the conformal curvature (the Schwarzian derivative of the Frenet curvature);  it is constant along equilibrium curves. 
Because tension is not a conformal invariant, conformal transformations of logarithmic spirals generally introduce tension and a length scale. All equilibrium states are generated this way. 
\end{abstract}

\today

\vskip 1em


\vskip 3em

\section{Introduction}

\begin{figure}[htb]
\begin{center}
\includegraphics[height=5cm]{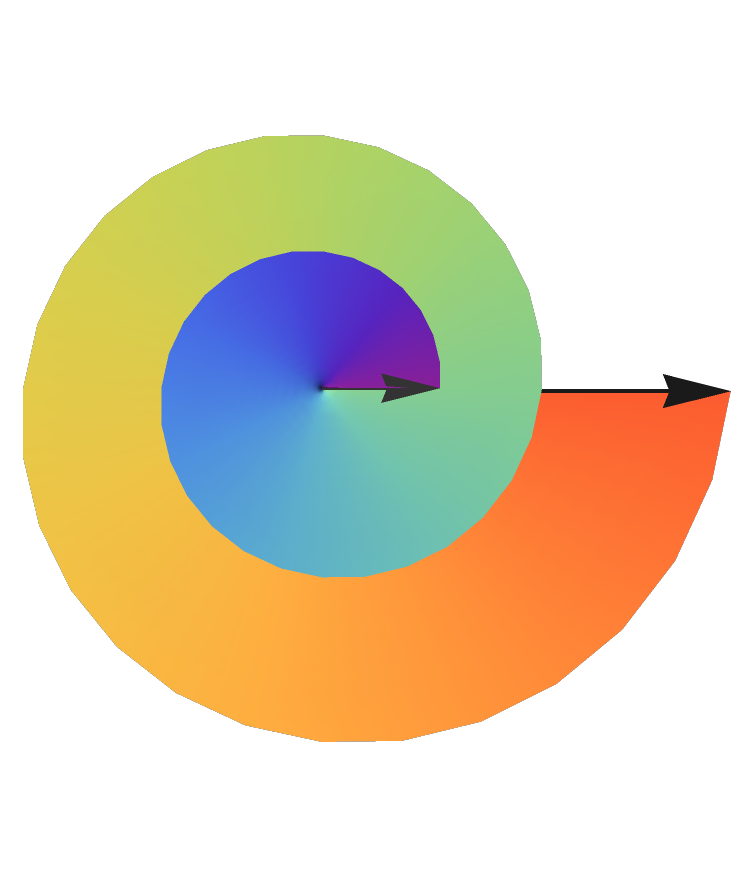}
\caption{\small \sf   Two revolutions of a logarithmic spiral.
As the spiral develops in an anti-clockwise fashion, the position vector from its apex sweeps out  an expanding multiply-covered region, its area 
proportional to  the square of the length of this vector. This is a direct consequence of the existence of a conserved 
special conformal current discussed in section \ref{Fzero}.}\label{Picture}
\end{center}
\end{figure}

Self-similar geometries are frequently encountered in the natural world. 
Perhaps the best-known and certainly the simplest examples are provided by logarithmic spirals.  They can be characterized in a number of ways. The simplest is a local statement: the  tangent at a point makes a constant angle with the radial direction. Indeed
the leading edge of a torn sheet of cellophane will spontaneously follow a log spiral, an observation that is not surprising on  one level, yet has significant implications \cite{Romero2013}. 
And, of course, anyone who has ever peeled an orange will  at some point have observed---with an element of satisfaction---that the peel can be nudged to follow 
a loxodrome, steering at a fixed angle to the meridian---the projection of a logarithmic spiral onto a sphere.  Equally well, a logarithmic spiral can be defined in terms of its growth: its radius of curvature increases linearly with the distance traveled along the curve.  
A logarithmic spiral famously also describes the cross-section of the nautilus shell, or the phyllotactic patterns in plants reflecting, as stressed by D'Arcy Thompson over a century ago, how the process of growth in biology very often generates self-similar structures \cite{Thompson}.  Nor is it a coincidence that this geometry is observed in the spiral arms in galaxies. Equivalently, a logarithmic spiral can be characterized by geometrical ratios: the distance traveled along the spiral path is proportional to the distance to the apex.  A perhaps even more fundamental non-local characterization is proposed in this paper:  it is the curve  whose distance to the apex is proportional to the nested multiply-covered planar area subtended by the radial vector, an identity conveyed in Figure \ref{Picture}. Each of these identities is a consequence of a conservation  law.  The apparently unlikely distance-area relationship, extraordinarily, turns out to be the one geometric definition that connects higher-dimensional generalizations of a logarithmic spiral with their planar prototype \cite{Paper3}:  the relevant area is the one projected orthogonal to the torque direction.
Each of these equivalent definitions of a logarithmic spiral implies its self-similarity: the curve is reproduced if scaled and rotated appropriately.  

\vskip1pc\noindent Logarithmic spiral crop up in the computer science literature (\cite{Marsland2016} for example). 
A logarithmic spiral is also conjectured to provide a solution to the  \textit{Search for the shore} problem posed by computer scientists: finding the \textit{optimal} trajectory to reach a straight line from a point on the plane when neither its distance nor its orientation is known.  It is clear that it must be a spiral;  as such intersecting all lines. Assuming reasonably that the solution will be self-similar, Finch et al. showed that it must be a logarithmic spiral \cite{Finch05,Finch2}. An interesting three-dimensional generalization is relevant when spatial analogues of logarithmic spirals are examined \cite{Paper3}. 
\\\\
In this paper, the simplest physical scenario in which self-similar curves are selected will be examined.
Whenever scale invariance is manifest in a physical system, invariably there 
is  a larger symmetry lurking. For curves on the plane, the relevant extension is the  
M\"obius group, consisting of similarities supplemented with inversions in circles and their compositions. 
From this point of view it is natural to construct an energy functional exhibiting this
invariance.\footnote{\sf Reflecting the relevance to curves in higher-dimensions, liberties will be taken using conformal invariance instead of the technically correct M\"obius invariance.} 
 The simplest, non-topological, scale invariant of a planar curve is its \textit{conformal} arc-length: 
\begin{equation}
H_0=  \int ds \, |\kappa'|^{1/2}\,,
\label{eq:H0}
\end{equation}
where $\kappa$ is the Frenet curvature along the curve and the prime denotes a derivative with respect to the Euclidean arc-length $s$.  
While $H_0$ is not an obvious energy in a Euclidean setting, 
it is the unique  M\"obius invariant at this order in derivatives. 
In contrast, scale invariants are many and easy to to easy to construct, among them the interpolation between $H_0$
and the  rotation number:
\begin{equation}
H_t=  \int ds \, (t\kappa'^2 + (1-t) \kappa^4)^{1/4}\,,\quad 0\le t\le 1\,.
\label{eq:S1}
\end{equation}
Unlike the conformally invariant $H_0$, the critical points of these more general energies are  not simple. 
\\\\
There are interesting discussions of $H_0$ in references \cite{Sharpe1994} and \cite{Fuster}.  The calculus of variations was first applied to Eq.(\ref{eq:H0})  by Liebmann \cite{Liebmann}; the subject has also been addressed,  more recently, in reference \cite{Bolt}. 
The critical points of $H_0$ are  logarithmic spirals and curves related to them by  conformal transformations which, unlike the logarithmic generator are not self-similar:  the compact double $\mathcal{S}$-shaped spirals and their hyperbolic limits, the latter arising when the center of inversion itself lies on the spiral.  
\\\\
Our approach to examining the critical behavior of $H_0$ will be to treat it as a mechanical energy,  mimicking as far as possible the approach to the Euler-Elastic  bending energy, quadratic in $\kappa$ \cite{SingerSantiago,Levien}. The reader is also referred to  more recent work in \cite{Hanna2018}.  Euler-Elastic curves  in three dimensions have also been the subject of intensive research; see, for example, \cite{LangerSinger,IveySinger}; enjoying an unexpected revival due to its tendency to crop up  in soft matter and biophysics in the description of semi-flexible polymers and thin filaments. In this context, Euler-Elastica  also plays a central role in the rapidly expanding of area of research,  dubbed  \textit{Extreme Mechanics} and has been adopted by at least one journal in its title. 
\\\\
The translational invariance of $H_0$ permits the Euler-Lagrange (EL) equations to be cast as a conservation law for 
the associated tension, $\mathbf{F}$, which will be identified as
\begin{equation}
\mathbf{F} =
 \frac{1}{2} \left[-
\big(\kappa\mu\big)'\,\, \mathbf{t}+ \mu''\, \, \mathbf{n} \right] \,.
\label{eq:Fkmu0}
\end{equation}
Here $\mathbf{t}$ and $\mathbf{n}$ are the vectors tangent and normal to the curve, and $\mu= |\kappa'|^{-1/2}$.
While $\mathbf{F}$ is not itself conformally invariant,
the EL equations $\mathbf{F}'=0$ are. They are also completely equivalent to the constancy of the conformally invariant curvature, 
\begin{equation}
\label{CalKmu}
\mathcal{K}= -\mu \left(\frac{d^2}{ds^2}  + \kappa^2/2\right) \mu  +  (\mu')^2/2\,.
\end{equation}
The torque, associated with rotational invariance,  is orthogonal to the plane and given by
\begin{equation}
\label{Mdef0}
M= (\mathbf {X} \times \mathbf{F})\cdot \mathbf{k}  + \mu'/2  \,,
\end{equation}
where the Cartesian description of the arc-length parametrized curve is given by the map,  $s\to \mathbf{X}(s)$, and $\mathbf{k}$ is a unit vector orthogonal to the plane. Noether's theorem can also be used to complete the  construction of the scaling and vector-valued special-conformal currents, $S$ and $\mathbf{G}$, associated with the additional conformal symmetry, 
\begin{subequations}
\label{GSdef}
\begin{eqnarray}
S &=& \mathbf{F}\cdot \mathbf{X} + \kappa \mu/2\,;\\
\mathbf{G} &=&
\mu' 
 \,  \mathbf{X}_{\perp}  -  |\mathbf{X}|^2 \,\mathbf{F}
 +  2 S  \, \mathbf{X}
-   \mu  \, \mathbf{n}\,,
\end{eqnarray}
\end{subequations}
where
\begin{equation}
\label{F0}
 \mathbf{X}_{\perp}=
(\mathbf{X}\cdot \mathbf{t})  \,\mathbf{n} - (\mathbf{X}\cdot\mathbf{n}) \, \mathbf{t} \,.
\end{equation}
Neither $S$ nor $\mathbf{G} $ possess Euler-Elastic counterparts. Notably, 
none of the currents are invariant under the conformal symmetry.  
\\\\
An intriguing property of scale invariant energies is that this invariance alone completely determines the tangential tension; reparametrization invariance then completely
fixes the normal tension, completing the construction of $\mathbf{F}$.  
\\\\
The tension vanishes in logarithmic spirals.   
The existence of non-trivial tension-free states is unusual in physics:  in contrast, the only tension-free open Euler-Elastic curves are straight lines. But this is not a coincidence:
the tension comes with a dimension.  
When it vanishes,  so also does $\mathbf{G}$ if referred to a specific point, identified as the spiral apex. 
As a consequence, the corresponding curve, charactered by the dimensionless $S$ and $M$, does not possess a length scale.  From this mechanical point of view, logarithmic spirals are self-similar because the tension within them vanishes. Indeed, they are the only self-similar stationary states  of the energy (\ref{eq:H0}) with $S$ and $M$ constrained to satisfy $4MS=1$, and as such characterized completely by  $S$, a measure of how rapidly the spiral unwinds. Its remarkable geometry notwithstanding, logarithmic spirals exhibit no internal structure.  
\\\\
The energy $H_0$ is very special in that the vanishing tension leads very directly  to the identification of 
all tension-free states as logarithmic spirals.
This happy accident is misleading: in general the  relationship between normal and tangential tension is not so direct.  The conservation laws associated with the conformal symmetry  provide a systematic method to construct tension-free states. This will become evident 
when  higher-order conformally invariant energies are contemplated. 
 It becomes even more strikingly evident when one examines the conformally invariant analogue of $H_0$ in higher dimensions in order to identify the spatial analogues of logarithmic  spirals \cite{Paper3}.  
 \\\\
The conformal curvature $\mathcal{K}$ is identified as the Casimir invariant  of the  
conformal group when the energy is given by $H_0$, a constant in equilibrium. Whereas $\mathbf{F}^2$ is invariant under the Euclidean group, it is \textit{not} under the conformal group.  It is straightforward to show that 
$\mathcal{K}$ is quadratic in the four conserved currents: 
$\mathcal{K}/2= \mathbf{F}\cdot \mathbf{G}-S^2 +  \mathbf{M}^2$. 
There is a logarithmic spiral for each constant value of $\mathcal{K}$. 
\\\\
The conformal invariance of the energy implies that the conformal descendants of a logarithmic spiral is also an equilibrium state with the same energy.  Tension will be generated in these states and they will not be self-similar. The pattern of tension generation and how it is reflected in the geometry is still of interest for it is not obvious what the connection is. The magnitude of the tension is, as one one expect, inversely proportional to the distance between the poles of these double spirals; but the tension is not simply directed along the line connecting the two poles,
reflects the chirality of its logarithmic progenitor.  
It be be confirmed that these states can also be 
constructed explicitly using the  four conserved currents to provide a  first order ordinary differential equation determining the trajectories of stationary states without appealing directly to  the conformal invariance of the system.  
\\\\
In general, the tension-free states associated with higher-order invariants  will  exhibit additional structure consistent with self-similarity. To illustrate this point, the conformally invariant energy, linear in $\mathcal{K}$, will be examined. Now the EL equation involves the constancy not of $\mathcal{K}$ but of  quadratic in derivatives of $\mathcal{K}$; the scaling current is linear in $\mathcal{K}$, whereas the torque is linear in its first derivatives. Significantly, $S$ and $M$ are no longer constrained in tension-free states. 
It will be shown how the two conservation laws can be combined to identify a quadrature for the dimensionless composite variable, $\kappa'/\kappa^2$.  The internal structure is described by the periodic oscillation of this variable with respect to the rotation angle about a logarithmic spiral. 
A framework is established for investigating how logarithmic spirals get decorated when the mechanics is controlled by higher order conformal invariants.  

\section{ Conformal invariant energies for planar curves}
\label{CIE}

Consider an arc-length parametrized curve $s\to \mathbf{X}(s)$ on the Euclidean plane with the  inner product between two vectors denoted by a centerdot separating them.
The unit  tangent vector to this curve is $\mathbf{t}=\mathbf{X}'$ where prime denotes a derivative with respect to arc-length,
and the its normal  $\mathbf{n}$ has cartesian components  $n^i=\epsilon^{ij} t^j$. Now  
\begin{equation}
\label{FS}
\mathbf{t}' = - \kappa \,\mathbf{n}\,;\quad
\mathbf{n}' =  \kappa \,\mathbf{t}\,,
\end{equation}
where $\kappa$  is the Frenet curvature. Modulo Euclidean motions, the curve is completely determined by $\kappa$. 
\\\\
Any Euclidean invariant energy can be expressed in terms of the Frenet curvature and its derivatives. Thus, without loss of generality, such an energy can be cast in the form
\begin{equation}
\label{Hdef}
H[ \mathbf{X}]=
\int ds\, \mathcal{H}(\kappa,\kappa', \dots )\,,
\end{equation}
where the ellipsis represents possible higher derivatives. 

\subsection{Conformal Invariants}
\label{CTs}

It is not hard to show that there is no non-trivial conformal  invariant energy constructed using $\kappa$ alone.\footnote{\sf
The invariant linear in curvature is the rotation number,  a topological invariant.} 
The simplest non-trivial conformal invariant involves $\kappa'$; indeed the only such invariant is proportional to the conformal  arc-length given by Eq.(\ref{eq:H0}).
To understand why, 
it is useful to possess a systematic approach to constructing invariants under conformal  transformations, the compositions of inversions in circles, Euclidean motions and scalings.  
The non-trivial element is inversion.   Thus consider inversion in a unit circle 
centered at the origin.\footnote{\sf The explicit use of complex variables is avoided in favor of a language that admits a generalization to higher dimensions.}
The point $\mathbf{X}$ on the curve maps to the point $\bar{\mathbf{X}} = \mathbf{X}/|\mathbf{X}|^2$;  
Now introduce 
the linear operator $\mathrm{R}_\mathbf{X}$ representing
a reflection in the line passing through the origin, orthogonal to (the unit vector) $\hat{\mathbf{X}}$,
\begin{equation}
\label{eq:Rdef}
\mathrm{R}_\mathbf{X} =
{\sf 1} - 2 \hat{ \mathbf{X}} \otimes \hat{\mathbf{X}}\,.
\end{equation}
The tangent and normal vectors transform 
${\mathbf{t}} \to |\mathbf{X}|^2 \mathrm{R}_\mathbf{X} \, \mathbf{t}
$, $\mathbf{n}\to -  |\mathbf{X}|^2 \mathrm{R}_\mathbf{X}\,\mathbf{n}$ (the latter with a minus sign). 
It follows that the arc-length 
transforms by
\begin{equation}
\label{arc-length}
ds \to d\bar s =ds/ |\mathbf{ X}|^2\,.
\end{equation}
As for the curvature,  it is relatively straightforward to show that  
$\kappa\to \bar \kappa $, where 
\begin{equation}
\label{eq:kapinversion}
\bar \kappa= -|\mathbf{ X}|^{2} \left( \kappa  - 2 \,(\mathbf{
X}\cdot \mathbf{ n})  / |\mathbf{ X}|^2\right)\,.
\end{equation} 
It is not a conformal scalar or primary field  (transforming homogeneously).
However, its first derivative, $\kappa'$, is a primary field with conformal weight $|\mathbf{ X}|^{4}$. This is because 
\begin{equation}
\frac{d\,\bar \kappa}{d{\bar s}}  = - |\mathbf{ X}|^2 \, \Big(|\mathbf{ X}|^{2} \kappa- 2 \,(\mathbf{
X}\cdot \mathbf{ n})  \Big)' 
= -|\mathbf{ X}|^4  \kappa' \,.\label{kprime}
\end{equation}
It follows as a consequence of Eqs.(\ref{arc-length}) and (\ref{kprime}) that the conformal arc-length $H_0[\mathbf{X}]$, defined by Eq.(\ref{eq:H0}), is a conformal invariant of planar curves.

\subsection{Conformal Curvature }
\label{HO}

The second and higher derivatives $\kappa''$  do not form primary fields. 
In section \ref{ConformalC}, it will be demonstrated that there are no local invariants involving $\kappa$ and its first two derivatives.
While the third derivative transforms in an even more complicated way, 
there is a conformal invariant involving $\kappa$ and its first three derivatives.
One can confirm that 
$4 \kappa'  \,\left(\frac{d^2}{ds^2} 
- \kappa^2 \right) \,\kappa'  - 5 \, {\kappa''}^2 
$ is a primary field transforming with conformal weight $|\mathbf{ X}|^{12}$ so that, on taking ratios,  
the conformal curvature,
\begin{equation}
\label{calKkappa}
\mathcal{K}=\frac{1}{ 8 {\kappa'}^{3}}\, \Bigg[4 \kappa'  \left(\frac{d^2}{ds^2} 
- \kappa^2 \right) \, \kappa'  - 5\, {\kappa''}^2 
\Bigg] \,,
 \end{equation}
 is identified  as a conformal scalar \cite{Sharpe1994}.
The simpler 
expression (\ref{CalKmu}) is reproduced if $\kappa'$ is eliminated in favor of the variable   
$\mu=|\kappa'|^{-1/2}$, transforming  (like arc-length) with weight $|\mathbf{X}|^{-2}$. Notice that the denominator appearing in Eq.(\ref{calKkappa}) is absorbed into the definition of $\mu$ and its derivatives. An important point is that 
$\mathcal{K}$ completely specifies the curve up to conformal transformations \cite{Sulanke1981}.
\\\\
In section \ref{ELS},  it will be demonstrated that  the EL derivative of the conformal arc-length is proportional to $\mathcal{K}'$. Higher order invariants will appear in the EL derivative of higher order energies such as the  conformal bending energy, 
\begin{equation}
\label{eq:HB}
H_B= \int d\mathcal{S} \, \mathcal{K}^2\,,
\end{equation}
where $d\mathcal{S}= ds\,\mu^{-1}$ is the conformal arc-length one-form.

\section{Conserved tension and critical points of curvature energies}
\label{aux}

The next task is to examine the response of conformal invariant energies to deformations of curves.  
Any two geometries related by a conformal transformation
will possess the same energy. Thus if one of them describes a critical point of this energy (so that it represents an equilibrium), then so does the other.  From the Euclidean point of view, in which the 
physics they represent is interpreted, the two curves can be  very different. In particular, the tension within the two will generally differ.  
\\\\
To derive the EL equation describing the equilibrium curves
the method of auxiliary variables will be useful \cite{auxil}.  While originally introduced to examine energies quadratic in curvature,  there is no obstacle to accommodating the conformal invariants described in section \ref{CIE}.  In this framework, the EL equation  is described in terms of the conservation of tension along these curves. From a conformal point of view, this is unusual: the EL equation is conformally invariant;  the tension is not. 
Of particular interest will be tension-free curves.  There is no length scale associated with 
such curves and they will form self-similar patterns.  For conformal arc-length these turn out to be logarithmic spirals; in general, additional internal structure makes an appearance.   
In contrast, there are no analogs of tension-free states of Euler-Elastica, unless trivially.
\\\\
To continue, let us examine the behavior of the
energy under small deformations. Let $\mathbf{X}\to\mathbf{X}+ \delta \mathbf{X}$. Modulo boundary terms, the change in the energy, $H[ \mathbf{X}]$, can always be cast in the form
\begin{equation}
\label{delHpperp}
\delta H[ \mathbf{X}] = \int ds\, \mathcal{E}_\perp\,\mathbf{n}
\cdot\delta\mathbf{X}\,.
\end{equation}
Critical points are characterized by the EL equation: $\mathcal{E}_\perp=0$. 
There is no corresponding tangential term $\approx \mathcal{E}_\| \,\mathbf{t}\cdot\delta\mathbf{X}$. This is a consequence of the reparametrization invariance of the energy
$H[ \mathbf{X}] $ and the identification of tangential deformations with reparametrizations at first order which together imply $\mathcal{E}_\|=0$. 
\\\\
Consider, more generally, any functional $H[ \mathbf{X}]$ defined on an arc-length  parametrized  curve that can be cast in the 
form (\ref{Hdef}).  The Frenet curvature $\kappa$ is constructed out of $\mathbf{X}$ and its first two derivatives. It is, however,  possible and very useful to treat it as an independent variable in its own right;  one way to do this consistently is to introduce the steps in its construction in terms of  $\mathbf{X}$ as constraints using Lagrange multipliers.  Thus the functional $H[\mathbf{X}]$ is replaced by 
$H_C[ \mathbf{X},\mathbf{t},\mathbf{n},\kappa,\dots]$, defined by 
 \begin{eqnarray}
H_C[ \mathbf{X},\mathbf{t},\mathbf{n},\kappa,\dots]  &=&  H[\kappa]
+ \int ds
\,\left[ \frac{1}{2} T (1 - \mathbf{t}\cdot \mathbf{t} )
- H_\kappa (\kappa- \mathbf{t}\cdot \mathbf{n}')\right]\nonumber\\
&&\quad\quad +  \int ds \left[ \frac{1}{2} \lambda (\mathbf{n}^2 -1) -  f\, (\mathbf{n}\cdot \mathbf{t})
+ \mathbf{F}\cdot (\mathbf{t}-
\mathbf{X}') \right]\,,\label{eq:aux}
\end{eqnarray}
in which $\mathbf{X},\mathbf{t},\mathbf{n}$ and $\kappa$ appear as independent variables.
The constraints that connect them are imposed by the Lagrange multipliers $T,$ $H_\kappa$, $\lambda$, $f$ and $\mathbf{F}$.
\\\\
The EL equations for $\mathbf{t}$ and $\mathbf{n}$ identify the tension $\mathbf{F}$ along the curve to be  given by 
\begin{equation}
\label{eq:Fdef}
\mathbf{F}=\big ( T - H_\kappa \kappa\big) \, \mathbf{t} - {H_\kappa}' \, \mathbf{n} \,.
\end{equation}
\\\\
The EL equations for $\kappa$ identifies $H_\kappa$ as the Euler Lagrange derivatives of the 
unconstrained functional $H[\kappa]$ with respect to 
$\kappa$:
\begin{equation}
H_\kappa= \frac{\delta H}{\delta \kappa}=
\frac{\partial \mathcal{H}}{\partial \kappa} -
 \left(\frac{\partial \mathcal{H}}{\partial \kappa'}\right)' +
\left(\frac{\partial \mathcal{H}}{\partial \kappa''}\right)'' - \left(\frac{\partial \mathcal{H}}{\partial \kappa'''}\right)'''
\,\,.\label{eq:Hdef}
\end{equation}
We admit a dependence on $\kappa$ and its first three derivatives. $H_\kappa$ does not vanish in
equilibrium.
\\\\
In this framework, the functions $\mathbf{X}$ appear only in the tangency constraint. 
$H_C$ is thus stationary  with respect to variations of $\mathbf{X}$ when 
\begin{equation}
\label{Fprime}
\mathbf{F}' =0\,, 
\end{equation}
or $\mathbf{F}$ is a constant vector along the curve.  The vector $\mathbf{F}$ determines the response of the energy to translations of the end points: as such it is identified as the tension in the curve and it will be  constant along equilibrium curves. By using the decomposition of the vector 
$\mathbf{F}$ along the tangent and normal,
 \begin{equation}
\label{Fdecomp}
\mathbf{F}= F_\| \,\mathbf{t} + F_{\perp} \,\mathbf{n}\,,
\end{equation}
where $F_\|$ and $F_\perp$ are given by Eq.(\ref{eq:Fdef}),
the vector-valued conservation law (\ref{Fprime})
can be rephrased as two scalar equations in terms of the projections of $\mathbf{F}$:
\begin{subequations}
\label{Fperppar}
\begin{eqnarray}
\mathcal{E}_\perp := F_\perp' - \kappa \, F_\| &=&0\,; \\
\mathcal{E}_\| := F_\|' + \kappa F_{\perp}  &=&0\,.
\end{eqnarray}
\end{subequations}
Eq.(\ref{Fperppar})(b)  implies that the normal tension $F_\perp$ is always completely 
determined by $F_\|$.
Modulo this equation, 
the EL derivative with respect to $\mathbf{X}$,
given by the right hand side of Eq.(\ref{Fperppar})(a), is cast completely in terms of $F_\|$:
\begin{equation}
\mathcal{E}_\perp= -\left[ \frac{1}{\kappa} F_\|' \right]'  - \kappa F_\|\,,
\label{EperpFpar}
\end{equation}
independent of the specific model.   
\\\\
It will be seen in section \ref{scaling}
that when $H$ is scale invariant,  not just the functional form of $F_\perp$ but also that of 
$F_\|$  itself is constrained.

\subsection{Reparametrization invariance and $\mathcal{E}_\|=0$}
\label{reparam}

The variational framework used so far has exploited the arc-length parametrization of the curve. There is a  price to pay: the breaking of manifest reparametrization invariance.  its consequence is that the tangential EL equation is no longer satisfied identically. Its role in this parameterization is to determine the 
multiplier $T$ imposing the unitary constraint on the tangent vector in Eq.(\ref{eq:aux}) which
flags $s$ as arc-length.
\\\\
To determine $T,$ the definition of $F_\|$ and $F_\perp$ in Eq.(\ref{eq:Fdef}) are used to recast Eq.(\ref{Fperppar})(b) in the form
\begin{equation}
\label{FparT}
(T  - 2 H_\kappa \kappa)'  +  H_\kappa \kappa'  =0\,.
\end{equation}
Using Eq.(\ref{eq:Hdef}) to eliminate $H_\kappa$, it follows that $T$ is given by
\begin{equation}
T  = - \mathcal{H}  +  2 H_\kappa  \kappa   +
P_\kappa \, \kappa'  +
P_{\kappa'} \, \kappa''  + P_{\kappa''} \kappa''' +
\cdots\,,
\label{eq:T1}
\end{equation}
where
the $P$s appearing in Eq.(\ref{eq:T1}) are given by  \footnote{\sf Only $P_\kappa$ is non-vanishing for conformal arc-length.  But a dependence on three derivatives needs to be accommodated  if energies involving the conformal curvature are contemplated.}
\begin{equation}
\label{3Ps}
 P_{\kappa} =
 \left(\frac{\partial \mathcal{H}}{\partial \kappa' }\right) -
 \left(\frac{\partial \mathcal{H}}{\partial \kappa''}\right)'  + 
  \left(\frac{\partial \mathcal{H}}{\partial \kappa'''}\right)'' \,; \quad
P_{\kappa'} = \left(\frac{\partial \mathcal{H}}{\partial \kappa''}\right)  - 
 \left(\frac{\partial \mathcal{H}}{\partial \kappa'''}\right)'  \,;\quad
P_{\kappa''} = \frac{\partial \mathcal{H}}{\partial \kappa'''}\,.
\end{equation} 
The tangential tension is now completely determined:
\begin{equation}
F_\|= T - H_\kappa \kappa  = - \mathcal{H}  +   H_\kappa \kappa  
+ P_\kappa \, \kappa' +  P_{\kappa'} \, \kappa'' + P_{\kappa''}\, \kappa'''\,.
\label{eq:Fpar1}
\end{equation}
It is perhaps surprising that the solution of   
the equation $\mathcal{E}_\|=0$ in favor of $T$ is completely tractable irrespective of the specific model  being examining.  Alternatively, and more explicitly, 
one could have evaluated $T$ directly imitating the treatment of surfaces \cite{auxil}; in this approach, the reparametrization invariance is manifest, and $T$ is identified as a one-dimensional  metric stress. 
Viewed this way, the identity (\ref{Fperppar})(b) is identically satisfied, the Bianchi identity associated with reparametrization invariance.
The two approaches end up, of course, in the same place. 
The identification of $T$ using the tangential EL equation in the approach adopted here restores the manifest reparametrization invariance that was temporarily suspended in choosing $s$ to parametrize the curve. 
\\\\
It will be seen in section \ref{scaling} that, when the energy is scale invariant, the steps taken in section (\ref{reparam}) are obviated; this is because the tangential stress is completely determined by the scaling current without any reference to the multiplier, $T$.

\section{Conformal arc-length and tension}
\label{ELS}

If $\mathcal{H}$ is a function only of $\kappa'$, the tangential tension (\ref{eq:Fpar1}) reduces to 
\begin{equation}
F_\| = - \mathcal{H}  +   H_\kappa \kappa  
+ \left(\frac{\partial \mathcal{H}}{\partial \kappa' }\right) -
   \, \kappa' \,.\label{Fpar2}
\end{equation}
In particular, if $\mathcal{H}(\kappa')={|\kappa'|}^{1/2}$, then $F_\|=
-\big(\kappa \mu  \big)'/2$, where  $\mu=\kappa'^{-1/2}$ introduced in section (\ref{HO}). Using
Eq.(\ref{Fperppar})(b)  to cast $F_\perp$ in terms of $F_\|$,  it follows that  
\begin{equation}
\mathbf{F} = \frac{1}{2}\, \left[-
\big( \kappa \mu\big)'\,\, \mathbf{t}+ \frac{1}{\kappa}\, \big( \kappa\mu\big)''\, \, \mathbf{n} \right] \,,
\label{eq:Fkmu}
\end{equation}
which coincides with Eq.(\ref{eq:Fkmu0}) on reorganizing the normal component.

\subsection{Tension-free states are logarithmic spirals}
\label{Logspiral}

A necessary and sufficient condition that $\mathbf{F}=0$ in equilibrium is that
$F_\|=0$.  This is because, modulo $F_\|=0$, 
Eq.(\ref{Fperppar})(b) implies that $F_\perp=0$, 
and Eq.(\ref{EperpFpar}) is satisfied identically.
\\\\
Using Eq.(\ref{eq:Fkmu}),  $F_\|=0$ is equivalent to the condition $(\kappa \mu)' = 0$, so that 
\begin{equation}
 \frac{\kappa^2}{|\kappa'|}=  4S^2\,,
\label{kprimek2}
\end{equation}
where $S$ is a constant which will be identified as the conserved scaling current in section \ref{scaling} (c.f. Eq.(\ref{GSdef}a)).  The curvature along spirals with positive $\kappa$ are
\begin{equation}
\label{kappa Ss}
\kappa= \frac{4 S^2}{s}\,,\end{equation}
which describes a 
logarithmic spiral. Thus tension-free curves are identified as logarithmic spirals. 
The mirror image has negative $\kappa$.
\\\\
A logarithmic spiral can also be defined, modulo a Euclidean motion, by the proportionality of the distance 
to the apex at $\mathbf{X}=0$ to the distance traveled along the curve,
\begin{equation}
\label{Xs}
|\mathbf{X}| =\cos \alpha\,\, s\,,
\end{equation} 
where $0\le\alpha\le\pi/2$ is constant. 
To establish the consistency with Eq.(\ref{kappa Ss}), 
first 
differentiate across the squared identity Eq.(\ref{Xs}) with respect to arc-length to obtain
 $\mathbf{t}\cdot \mathbf{X} = \cos^2\alpha\,s$, reproducing the familiar  result that the tangent vector  meets the radial vector at the constant angle, $\alpha$.\footnote{\sf In polar coordinates, $\rho=\cos\alpha\, s$. Using the unit vector identity $\rho'{}^2+ \rho^2 \phi'{}^2=1$,  implies  
$\phi = 4S^2 \,\ln s$.  $\phi$ thus coincides with the rotation angle $\Theta$, defined by $\Theta=\int ds \,\kappa$. }\label{Thetaphi}
 Differentiating Eq.(\ref{Xs}) once again implies $\kappa \mathbf{n}\cdot \mathbf{X} = \sin^2 \alpha$.
 The completeness of $\mathbf{t}$ and $\mathbf{n}$ then implies
$\cos^2\alpha  \,\, s^2 = \sin^2\alpha /\kappa^2$,
which coincides with the expression   (\ref{kprimek2}) for $\kappa$, with the identification
\begin{equation}
\label{Salpha}
2 S   = \tan\alpha  \quad{\sf or}\,\quad  \cos^2\alpha= 1/(1+ 16 S^4)\,. 
\end{equation}
\\\\
The constant conformal curvature in a logarithmic spiral can be determined in terms of $S$ using Eq.(\ref{CalKmu}):
\begin{equation}
\label{calCS}
\mathcal{K}_0 = \frac{1-16 S^4 }{ 8S^2}\,.
\end{equation} 
In particular, $\mathcal{K}_0=0$ if and only if $S= 1/2$. Logarithmic spirals with different values of $S^2$ are conformally inequivalent.  Similarity transformations do not map one into another.  

\subsection{$\mathcal{K}$ is constant in equilibrium:}
\label{calKC0}

The EL derivative of $H_0$ with respect to $\mathbf{X}$ is directly related to the conformal curvature, $\mathcal{K}$.  Substituting the 
expression for $\mathbf{F}$ given on the second line in Eq.(\ref{eq:Fkmu0}) into Eq.(\ref{EperpFpar}), the identity 
\begin{eqnarray}
2 \mu\, \mathcal{E}_\perp &=& \mu \mu''' + \kappa \mu (\kappa \mu)'
=\left(\mu \mu'' -\mu'^2/2 + (\kappa \mu)^2/2\right)' 
\end{eqnarray}
follows.  The definition of  $\mathcal{K}$ in terms of $\mu$ given by (\ref{CalKmu}), completes 
 the identity \cite{obs}
\begin{equation} \mathcal{E}_\perp =-\frac{1}{2\mu}  \,\mathcal{K}'\,.
\label{EperpK}
\end{equation}
The EL derivative is proportional to the conformal invariant, $d \mathcal{K}/d\mathcal{S}$, where 
$d/d\mathcal{S}= \mu d/ds$ is the derivative with respect to conformal arc-length.  Thus whereas 
$\mathbf{F}$ itself is not a conformal invariant, $\mathbf{F}'$ is. 
In particular,
the  EL equation, $\mathcal{E}_\perp=0$, and the conservation law,  $\mathbf{F}'=0$, are equivalent conformally invariant statements.   
In equilibrium, $\mathcal{K}$ is constant. 
The solutions of the EL equation  are thus partitioned into conformal equivalence classes, characterized by this constant, each of which contains a tension-free logarithmic spiral, characterized by the scaling rate $S$. \\\\
Had ones interest been limited to identifying the tension-free states of $H_0$ 
this would have been a short short, concluded at section \ref{Logspiral}. In general, however, when higher order conformal invariants are examined, the identity $F_\|=0$ is not sufficient to determine the curvature, 
never mind characterize the internal structure of the trajectories represented by this curvature. Even for conformal arc-length, it is worthwhile examining the conservation laws to understand how conformal  
invariance constrains trajectories.

\section{Boundary Variations and Conservation Laws}
\label{Conservationlaws}

The conserved currents associated with the symmetries of the energy
can be identified by examining the boundary terms accumulated in the variation of 
$H_C$, defined by Eq.(\ref{eq:aux}).
Restoring all  boundary terms, one finds that 
\begin{equation}
\label{eq:delHQ0def}
\delta H_C[ \mathbf{X},\dots]  =   \int ds\, \mathbf{F}' \cdot \delta\mathbf{X} +  \int ds\, \mathcal{Q}'\,,
 \end{equation}
 where
 \begin{eqnarray}
\mathcal{Q}&=&  H_\kappa \, \mathbf{t} \cdot \delta \mathbf{n}
 -  \mathbf{F}\cdot \delta 
\mathbf{X} 
 + P_{\kappa}\, \delta \kappa  + P_{\kappa'} 
\, \delta (\kappa') + P_{\kappa''} 
\, \delta (\kappa'') 
 \,,\label{eq:Qdef}
\end{eqnarray}
where $H_\kappa$ is the unconstrained EL derivative with respect to $\kappa$ given by
Eq.(\ref{eq:Hdef})
and $P_{\kappa}$, $P_{\kappa'}$ and $P_{\kappa''}$ are defined in Eq.
(\ref{3Ps}).
The first two terms originate in the variations of $H_C$  with respect to $\mathbf{X}$ and $\mathbf{n}$ when derivatives are peeled off the variation and collected in a derivative.  For the familiar Euler-Elastic energy, or any energy involving $\kappa$ but not its derivatives,  all $P$s vanish, and these are the only two boundary terms. If $H$ depends on first (second or third) derivatives of $\kappa$, additional terms make an appearance.  For conformal arc-length only 
$P_{\kappa}$ appears. If, however, $\mathcal{H}$ involves the conformal curvature
it will depend on $\kappa,\kappa',\kappa''$ and $\kappa'''$ and $P_{\kappa'}$ and $P_{\kappa''}$ will also be non-vanishing.  
On stationary curves, the first term in Eq.(\ref{eq:delHQ0def}) vanishes,
and only the boundary terms survive.

\subsection{Rotational Invariance and Torque}
\label{Torque}

Under a rotation  of the plane by an angle $\omega$ about the origin, to first order $\omega$, 
\begin{eqnarray}
\delta_\omega \, H_C
  &=& \omega \,\left( -\int ds \, (\mathbf {X} \times \mathbf{F}')\cdot \mathbf{k}  +  \int ds\, 
M' \right)\,, \label{eq:cb}
\end{eqnarray}
where 
\begin{equation}
\label{Mdef}
M= (\mathbf {X} \times \mathbf{F})\cdot \mathbf{k}  - H_\kappa  \,.
\end{equation}
In equilibrium, with $\mathbf{F}'=0$, the torque $M$ is conserved, $M'=0$.
If  $\mathbf{F}$ is directed along $\mathbf{i}$, and $\mathbf{X}=(X,Y)$, then for the conformal arc-length, 
\begin{equation}
M=\mu'/2 -  F Y\,,
\label{MkFi}
\end{equation}
where $\mu=|\kappa'|^{-1/2}$, which is equivalent to Eq.(\ref{Mdef0}).
\\\\
If $F=0$, $M$ is given in terms of $S$, the constant
appearing in Eq.(\ref{kappa Ss}), by the relationship $4M S= 1$.
Thus the torque and the scaling current $S$ (to be constructed in section \ref{scaling}), are not independent in logarithmic spirals. In tension-free states this is a Euclidean invariant statement. indeed the definitions of neither $M$ nor $S$ are translationally invariant if $\mathbf{F}\ne 0$.  This ambiguity is addressed in section \ref{GeneratingTension}. 
In general, for higher order conformally invariant energies, the is no simple relationship connection between $M$ and $S$ in tension-free states.   

\section{Conformal
 Invariance and its manifestations}

Let us now re-focus the discussion on conformally invariant energies, and the identification of the currents capturing this additional invariance.  Here it will become instructive to examine these currents not only in equilibrium but also away from it.   
First let us look at scaling. As we will see, the behavior of a scale invariant energy   implies a constraint on the form of $F_\|$, without reference to equilibrium.  This has far-reaching consequences.

\subsection{Scaling Current}
\label{scaling}

Under a rescaling, $\delta_\nu \mathbf{X} =\nu \mathbf{X}$,  $\delta_\nu \mathbf{n} = 0$, whereas 
$\delta_\nu \kappa = - \nu \kappa$, 
$\delta_\nu \kappa'= -2 \nu \kappa'$ and so on, reflecting the dimensionality of each variable. 
Substituting into Eq.(\ref{eq:Qdef}), 
we set $\mathcal{Q}= -\nu S$, where the scaling current $S$ is given by
\begin{equation}
\label{eq:Scaledef}
 S=  \mathbf{F}\cdot 
\mathbf{X} 
+ S_D\,,\end{equation}
and 
\begin{eqnarray}
\label{eqLSDdefN2}
S_D &:=&  P_{\kappa} \,  \kappa + 2 \, P_{\kappa'} \, \kappa'  +
3 P_{\kappa''} \,\kappa''  
 \,,
\end{eqnarray}
where $P_{\kappa}$, $P_{\kappa'}$ and $P_{\kappa''}$ are defined by
Eq.(\ref{3Ps}). 
Note the factors of two and three multiplying the second and third terms in $S_D$.
The current $S$ satisfies
\begin{equation}
\label{eq:Sprime}
S' = \mathbf{F}'\cdot \mathbf{X}
\end{equation}
whenever the energy is scale invariant. 
$S$ is thus conserved if $\mathbf{F}$ is.
\\\\
If $\mathcal{H}=\mathcal{H}(\kappa)$, without derivatives, then $S_D=0$ and Eq.(\ref{eq:Sprime}) implies that $F_\|=0$ if $H$ is scale invariant.  But,  as seen in section \ref{aux}, this would imply that $\mathbf{F}$ vanishes identically. There thus can be no non-trivial scale invariants constructed with $\kappa$ alone.
More generally, Eq.(\ref{eq:Sprime}) is equivalent to the statement that 
\begin{equation}
\label{Fparaderiv}
F_\| = - S_D' \,.
\end{equation}
Thus, whenever $H$ is scale invariant (and whether the curve is in equilibrium or not), $F_\|$ is expressible as a derivative. 
This simple identify is very useful 
bypassing as it does the need to evaluate the multiplier $T$ appearing in the definition (\ref{eq:aux}) of $H_C$ whenever the energy is scale invariant.  It is especially useful to bear in mind when higher order energies are being considered. 
We have already encountered  this identity for the conformal arc-length, by simply collecting terms. It would be reasonable to say, however, that  the identity (\ref{Fparaderiv}) is not evident in the construction of $F_\|$ leading to Eq.(\ref{eq:Fpar1}).  There is now an explanation. 
Note that in a scale invariant theory, the normal EL derivative (\ref{EperpFpar}) can now be cast in the form 
\begin{equation}
\mathcal{E}_\perp= \left[ \frac{1}{\kappa} S_D''  \right]'  + \kappa S_D'\,.
\label{ELSD}
\end{equation}
The identity (\ref{Fparaderiv})
is, of course, the Euler scaling equation associated with any scale invariant energy, albeit in disguise.
For, if  
$\mathcal{H}=\mathcal{H}(\kappa,\kappa',\kappa'',\kappa''')$,
Eqs.(\ref{eq:Fpar1}) and (\ref{Fparaderiv}) together imply
\begin{equation}
 - \mathcal{H}  +   H_\kappa \kappa  
+ P_\kappa \, \kappa' +  P_{\kappa'} \, \kappa'' + P_{\kappa''}\, \kappa''' =
- (P_{\kappa} \,  \kappa + 2 \, P_{\kappa'} \, \kappa'  +
3 P_{\kappa''} \,\kappa'')' 
\,.
\label{eq:scale21}
\end{equation}
Disassembling the $P's$ defined in Eq.
(\ref{3Ps}) and $H_\kappa$ (Eq.
(\ref{eq:Hdef})),  into their constituent partial derivatives and 
collecting terms, the identity 
\begin{equation}
  - \mathcal{H}  +   \left( \frac{\partial \mathcal{H}}{\partial \kappa}\right)\, \kappa  +
2 \left( \frac{\partial \mathcal{H}}{\partial \kappa'}\right) \,  \kappa'  + 
3 \left( \frac{\partial \mathcal{H}}{\partial \kappa''}\right) \,  \kappa''  + 
4 \left( \frac{\partial \mathcal{H}}{\partial \kappa'''}\right) \,  \kappa''' 
  = 0\,,
\label{Euler}
\end{equation}
follows.
There is, of course,  enormous liberty constructing scale invariants involving $\kappa$ and $\kappa'$. A non-trivial one-parameter family was written down in Eq.(\ref{eq:S1}).  It is, however, now clear that, the only possibilities involving first derivatives are of the form
$\kappa \mathcal{F}(\kappa'/\kappa^2)$, where $\mathcal{F}$ is some function of its argument.   The only conformally invariant among them, however,  is the conformal arc-length $H_0$, with $\mathcal{F}(x)=(x^2+1)^{1/4}$. There are even more scale invariants if $\kappa''$ is admitted, but not a single new conformal invariant. As was pointed out in section \ref{HO},  one needs to proceed to a dependence on the conformal curvature involving three derivatives. This issue will  be re-examined 
in section \ref{identifyconf} by examining the conservation of the special conformal current.
\\\\
For the conformal arc-length,  with $\mathbf{F}$ aligned along $\mathbf{i}$, $S= FX + \kappa \mu/2$, reproducing 
Eq.(\ref{GSdef})(a).  Notice that the scaling current is not a Euclidean invariant unless $F=0$. Indeed, if $F\ne 0$, it is possible to set $S=0$ using an appropriate translation parallel to  $\mathbf{F}$.

\subsection{Special Conformal Current}
\label{ConformalC}

To identify the special conformal current first examine the behavior of a curve under special conformal transformations, the composition of an inversion with a translation followed by a second inversion, lineared in the intermediate translation $\delta \mathbf{c}$: this takes the form
$\delta_\mathbf{c} \mathbf{X} = |\mathbf{X}|^ 2 \, \mathrm{R}_\mathbf{X} \, \delta \mathbf{c}$,
where $\mathrm{R}_\mathbf{X}$ is the linear operator defined in Eq.(\ref{eq:Rdef}). The constant vector $\delta \mathbf{c}$ has dimensions
of inverse length squared.
\\\\
One finds that
\begin{equation}
\label{eq:dsc}
\delta _\mathbf{c}  ds  =  -2\,  (\mathbf{X}\cdot\mathbf{c})\, ds \,.
\end{equation}
In addition, 
\begin{equation}
\label{dnc}
\delta_\mathbf{c} \mathbf{n} = 
2 (\mathbf{X}\cdot\mathbf{n}) \,\mathbf{c} - 2 (\mathbf{n}\cdot\mathbf{c})\, \mathbf{X}\,,
\end{equation}
and 
\begin{subequations}
\label{dkappaprime012}
\begin{eqnarray}
\delta_\mathbf{c}  \kappa &=& 2 [ (\mathbf{X}\cdot\mathbf{c})\, \kappa - (\mathbf{n}\cdot\mathbf{c})]\,;\\
\delta_\mathbf{c} \kappa'  &=& 4 (\mathbf{X}\cdot\mathbf{c})\,\kappa' \,; \\
 \delta_\mathbf{c}  \kappa'' &=&  6 \,(\mathbf{X}\cdot\mathbf{c})\, \kappa'' + 4
(\mathbf{t}\cdot\mathbf{c})\, \kappa' 
\,.
 \end{eqnarray}
\end{subequations}
Using Eq.(\ref{eq:Qdef}) with $
\mathcal{Q}=  \mathbf{G}\cdot \mathbf{c}$, and the results just collected,  the conformal current $\mathbf{G}$ associated with any conformally invariant energy  involving three derivatives or less of $\kappa$ is identified: 
\begin{equation}
\mathbf{G}= - 2 H_\kappa \,  \mathbf{X}_{\perp}  -   |\mathbf{X}|^2 \mathbf{F} +  2 S\,  \mathbf{X}    
-2 \,P_{\kappa}\, \mathbf{n} + 4\,P_{\kappa''}\,\kappa'\, \mathbf{t}
 \,.\label{eq:Gdef}
\end{equation}
The variables $H_\kappa$, $P_{\kappa}$ and $P_{\kappa''}$ are defined explicitly in Eqs.
(\ref{eq:Hdef}) and (\ref{3Ps}). 
The identity  
$-   |\mathbf{X}|^2 \mathrm{R}_\mathbf{X} \mathbf{F} +  2 S_D\,  \mathbf{X}  =
-   \mathbf{F} +  2 S\,  \mathbf{X}$,
as well as the decomposition of $S$ given by Eqs.(\ref{eq:Scaledef}) and (\ref{eqLSDdefN2}) have been used to simplify. Note that $\mathbf{X}_{\perp}$, defined by Eq.(\ref{F0}) in the introduction, is orthogonal to $\mathbf{X}$, with equal magnitude, $ |\mathbf{X}_{\perp}|^2=|\mathbf{X}|^2$. In addition,    
\begin{equation}
\mathbf{X}_{\perp}' = \mathbf{n} \,,
\label{eq:DF0}
\end{equation}
which identifies $\mathbf{X}_{\perp}$ as a potential for the normal vector \cite{Laplace},
$\mathbf{X}_{\perp}=(Y,-X)$. 

\subsection{The Casimir Invariant and Conserved Currents}
\label{Casimirsect}

For the conformal arc-length the conformal current 
$\mathbf{G}$ given by Eq.(\ref{eq:Gdef}) simplifies. Using the 
identity $P_\kappa= \mu /2$,  
Eq.(\ref{GSdef})(b) is reproduced.
\\\\
The isomorphism 
between the conformal group and the four-dimensional Lorentz group 
implies the identity between the conformal curvature and a quadratic in the 
currents: 
\begin{equation}
\label{Casimir}
\mathcal{K}/2= \mathbf{G}\cdot \mathbf{F}- S^2 + M^2\,.
\end{equation}
The individual currents $\mathbf{F},\mathbf{G}, M$ and $S$ are not invariant; but $\mathcal{K}$ is. $\mathcal{K}$ is identified as the Casimir  invariant of the conformal group for the conformal arc-length. When the four currents are conserved $\mathcal{K}$ is also.
In a more general conformal invariant theory, as demonstrated in section \ref{linKappa},
 the Casimir invariant will not coincide with the conformal curvature, but instead will be some higher order conformal invariant;  nor will $\mathcal{K}$ generally be conserved.

\section{Trajectories from the conformal current }
\label{Stationarystates}

First examine the special conformal current in tension-free states. These are necessarily logarithmic spirals so the reader would justifiably not expect to gain any additional insight by studying the conservation of $\mathbf{G}$
when $\mathbf{F}=0$. However,  even if this were the case, it is not true of higher order invariant energies or indeed for conformal arc-length when one steps up a dimension \cite{Paper3}; the conservation law has unexpected consequences even for conformal arc-length. 

\subsection{$\mathbf{F}=0$}
\label{Fzero}

If $\mathbf{F}=0$, then the definition of the torque given by Eq.(\ref{Mdef}) implies $H_\kappa$ is constant, $H_\kappa=-M$.
Thus under translation, the conformal current in a tension-free state transforms  by a constant vector:
\begin{equation}
 \mathbf{G}(\mathbf{X}+ \mathbf{a}) =  \mathbf{G}(\mathbf{X})  + 2 M 
  [(\mathbf{a}\cdot \mathbf{t})  \,\mathbf{n}- (\mathbf{a}\cdot\mathbf{n}) \, \mathbf{t}] 
+ 2 S \,\mathbf{a}\,.
\label{Gtranslate}
\end{equation}
This invariance can be exploited to translate the  tension-free curve so that $\mathbf{G}=0$  by appropriately choosing $\mathbf{a}$.  This places the apex at the origin.  As Eq.(\ref{Gtranslate}) indicates, this is generally true for the tension-free states of any conformally invariant energy. This can be understood from a Lie-group theoretical framework: $\mathbf{F}$ is the generator of translations; whereas $\mathbf{G}$ generated translations modulo inversion, a duality manifest in the Lie algebra of the conformal group (see, for example, reference \cite{NakayamaReview}). 
\\\\
In a tension-free state of the conformal arc-length, with $F_\|=0$, the scaling identity (\ref{Fparaderiv}) implies $\kappa \mu /2 =S$, where $S$ is the conserved scaling current reproducing Eq.(\ref{kprimek2}) and its 
 solution, $\kappa=  4 S^2/ s$,  describing a logarithmic spiral.  The vanishing conserved current provides 
 an alternative derivation of the connection between $\kappa$ and $|\mathbf{X}|$ discussed in section \ref{Logspiral}. To see this, note that 
when $\mathbf{G}=0$  and $\mathbf{F}=0$,  Eq.(\ref{GSdef})(b)
 implies  
\begin{equation}
0 =     -\frac{1}{2 S }  \,  \mathbf{X}_{\perp}
+ 2 S\,\mathbf{X} +   \frac{s}{2S}\,\mathbf{n}\,,
\label{eq:GF0}
\end{equation}
where $\mathbf{X}_{\perp}$ is given by Eq.(\ref{F0}). 
Projecting Eq.(\ref{eq:GF0}) onto the orthogonal vectors,
$\mathbf{X}$ and $\mathbf{X}_{\perp}$, one obtains
 \begin{eqnarray}
4 S^2\,|\mathbf{X}|^2 + s\,(\mathbf{X}\cdot\mathbf{n})&=&0\nonumber\\
-  \,  |\mathbf{X}|^2
 + s\,(\mathbf{X}\cdot\mathbf{t}) &=&0\,.
\end{eqnarray}
These two equations imply 
\begin{equation}
(1+ 16 S^4)  |\mathbf{X}|^2 = s^2\,,
\end{equation}
which, with the identification (\ref{Salpha}), is the spatial description of a logarithmic spiral treated in section \ref{Logspiral}. 
Choosing the origin so that $\mathbf{G}=0$ places it at the apex of the logarithmic spiral. 
\\\\
The identity, illustrated in Figure \ref{Picture},  
is obtained by projection of Eq.(\ref{eq:GF0}) onto the tangent direction, giving
\begin{equation} 
 \, (\mathbf{X}\cdot \mathbf{n})  
 =  2S^2  \, (|\mathbf{X}|^2)'\,,
 \end{equation}
 or 
\begin{equation}
 S^2   \, |\mathbf{X}|^2 = \frac{1}{2} \, \int ds \, (\mathbf{X}\cdot \mathbf{n})  =  A\,,
 \label{eq:rhoA}
\end{equation}
where $A$ is the area of the multiply covered region illustrated in Figure 1 that is generated  by the spiral as it unwinds.\footnote{\sf Using the results of section \ref{Logspiral}, 
 $(\mathbf{X}\cdot \mathbf{n})^2= |\mathbf{X}|^2 (1- \alpha^2)$, so that 
$4A= \alpha (1- \alpha^2)^{1/2} s^2$. Eq.(\ref{eq:rhoA}) then implies $S^2= (1- \alpha^2)^{1/2} /(4\alpha)$ which reproduces Eq.(\ref{Salpha}.}

\section{A  note on $\mathbf{G}$ and the identification of conformal invariants}
\label{identifyconf}

Using Eq.(\ref{eq:delHQ0def}), it is evident that 
the current 
$\mathbf{G}$  associated with any conformally invariant energy satisfies 
\begin{eqnarray}
 \mathbf{G}' &=& -   |\mathbf{X}|^2 \mathrm{R}_\mathbf{X} \mathbf{F}'\nonumber\\
 &=& 2 \mathbf{F}'\cdot \mathbf{X}\, \mathbf{X}  
  -   |\mathbf{X}|^2 \mathbf{F}'\nonumber\\
 &=& 2 S'\,  \mathbf{X} 
  -   |\mathbf{X}|^2 \mathbf{F}'\,.
  \label{eq:Gprime}
\end{eqnarray}
It is thus clear that $\mathbf{G}$ is conserved when $\mathbf{F}$ (and along with it $S$) is.  
Eq.(\ref{eq:Gprime}) together with  Eq.(\ref{eq:Sprime}) constraint the functional form of a Euclidean invariant energy if it is to be conformally invariant.
Eq.(\ref{eq:Gprime}) 
can be cast as  the following pair of equations, corresponding to its projections 
on $\mathbf{n}$ and $\mathbf{t}$ respectively:
\begin{subequations} 
\label{CIconst}
\begin{eqnarray} 
2\,
\left(\frac{\partial \mathcal{H}}{\partial \kappa''' }\right) \,\kappa\kappa'
+ \left (\frac{\partial \mathcal{H}}{\partial \kappa }\right)  &=&0 \,;\\
2 
\left (\frac{\partial \mathcal{H}}{\partial \kappa'' }\right)\,\kappa' 
+ 5\left (\frac{\partial \mathcal{H}}{\partial \kappa''' }\right) \kappa'' 
&=&0\,. 
\end{eqnarray}
\end{subequations} 
Curiously, these identities do not involve first derivatives.
They are thus automatically satisfied by the conformal arc-length. If $H$ depends on $\kappa'$ alone,
the appropriate power of $\kappa'$ is fixed by scale invariance, also captured in the scaling identity, Eq.(\ref{Euler}). 
 Nor do there exist any additional conformal invariants involving $\kappa,\kappa'$ and $\kappa''$. As argued in section \ref{HO},  it is  necessary to proceed to third order (accommodating $\kappa'''$) to construct 
an invariant other than $S$. 
\\\\
Eq.(\ref{CIconst})(a) implies that $H$ cannot depend on $\kappa$ unless it also depends on $\kappa'''\,$\footnote{
In this context, note that the topologically invariant rotation number involves boundary terms. Our search strategy only discovers strict conformal invariants. An interesting challenge is to accommodate boundary terms.};
Eq.(\ref{CIconst})(b) implies that $H$ also cannot depend on $\kappa''$ unless it also depends on $\kappa'''$ or higher derivatives.  After conformal arc-length, the next simplest invariant of a planar curve is a function of the conformal curvature, involving $\kappa$ and its first three derivatives. If $\mathcal{H}=\mathcal{F} (\kappa', \mathcal{K})$, then 
$\mathcal{K}$ also satisfies  Eqs.(\ref{CIconst}). In particular,
$\mathcal{K}= ( 4 \kappa' (\kappa''' -\kappa^2 \kappa')
 - 5 \kappa''{}^2)/8\kappa'{}^3$ 
satisfies Eq.(\ref{CIconst}). The normalization by a function of 
$\kappa'$, to reproduce the definition of the conformal curvature 
is fixed by scale invariance.

\section{Integrating the conservation laws when  $\mathbf{F}\ne 0$}

All equilibrium states arise 
as conformal descendants of tension-free  states. And this provides a direct method to construct these states.  This approach is also the most efficient and will be addressed in section \ref{GeneratingTension}. Additional insight can, however, be obtained by examining how the conservation laws
can be used to construct these states. 

\subsection{Constant $\mathbf{F}$}
\label{IntegrateF}

First examine the conservation law $\mathbf{F}'=0$.  This reproduces the conservation for $S$ and $M$ discussed in subsection \ref{scaling} and \ref{Torque} respectively.
\\\\
To see this, let the constant equilibrium tension be directed along the vector $\mathbf{i}$: $\mathbf{F}= F \mathbf{i}$.  The tangent and normal vector can be expressed in terms of their projections along the orthogonal vectors $\mathbf{i}$ and $\mathbf{j}$,  $\mathbf{t}= (X',Y')$ and $\mathbf{n}= (Y',-X')$).  Now, upon integration, the projection of $\mathbf{F}$, given by Eq.(\ref{eq:Fkmu0}), onto $\mathbf{t}$ implies  
\begin{equation}
2 (F X +   M_1)   = - \kappa \mu \,,
\label{FX}
\end{equation}
where $M_1$ is a constant of integration. As we will show,
Eq.(\ref{FX}) reproduces the conservation of the scaling current Eq.(\ref{eq:Scaledef}) discussed in subsection \ref{scaling},  with the identification 
$S= -M_1$ follows.  
The projection onto $\mathbf{n}$ gives
\begin{equation}
2 (F Y + M_2)   = \mu'\,,
\label{FY}
\end{equation}
where $M_2$ is a second constant of integration.  Eq.(\ref{FY}) reproduces the conservation of torque Eq.(\ref{MkFi}) discussed in subsection \ref{Torque} with the identification, $M_2=M$. 
If $F\ne 0$, it is always possible to perform a  translation along $\mathbf{i}$ so that the point where $\kappa=0$ lies at the origin and $M_1=0$. 
The conservation of the tension subsumes the conservation laws associated with scaling and rotations.  To complete the construction of 
equilibrium trajectories,  we now appeal to the additional input implied by the conserved conformal current, introduced in section \ref{ConformalC}.  

\subsection{Constant $\mathbf{G}$}
\label{IntegrateG}

When  $\mathbf{F}\ne0$, the conformal current $\mathbf{G}$, given by Eq.(\ref{GSdef})(b), cannot be translated away.  Its projections along $\mathbf{X}$ and $\mathbf{X}_{\perp}=(Y,-X)$  give respectively
\begin{subequations}
\label{GX0}
\begin{eqnarray}
\mathbf{G}\cdot \mathbf{X} &=&   |\mathbf{X}|^2 \,( 2S   -   \mathbf{F}\cdot \mathbf{X} )
 -   \mu\, (\mathbf{n}\cdot \mathbf{X} )\,;\\
\mathbf{G}\cdot \mathbf{X}_{\perp} &=& \mu' \, |\mathbf{X}|^2    -  |\mathbf{X}|^2  \,\mathbf{F}\cdot \mathbf{X}_{\perp}
-   \mu\,  { |\mathbf{X}|^2}'/2 \,.
\end{eqnarray}
\end{subequations}
Using  Eqs.(\ref{FX}) and (\ref{FY}) to eliminate $\mu$ and $\mu'$ in favor of $X$, $Y$ and $\kappa$, 
as well as the identity $M_1=-S$, Eqs.(\ref{GX0}) read
\begin{subequations}
\label{GX1}
\begin{eqnarray}
G_1 X + G_2 Y &=&   - |\mathbf{X}|^2 \, (2M_1 + FX) 
+  2 (F X +   M_1)\,  \frac{1}{\kappa} \,  (Y' X - X' Y)\,;\\
G_1 Y  - G_2 X 
&=& (F Y + 2 M_2) \, |\mathbf{X}|^2  + 2 (F X +   M_1) \,\frac{1}{\kappa} (XX' + YY')  \,,
\end{eqnarray}
\end{subequations}
or, equivalently, taking appropriate linear combinations:
\begin{subequations}
\label{GYXprime}
\begin{eqnarray}
2 F (F X +   M_1)\,   \frac{1}{\kappa} \,  Y' &=&
 ( FX+ M_1)^2- (F Y +  M_2)^2   +  G_1 F  - M_1^2 + M_2^2\\
2  (F X +   M_1)\, \frac{1}{\kappa} \,   X' 
& =&  - \Big( 2FXY +2 (M_1 Y + M_2 X) + G_2\Big)\,.
\end{eqnarray}
\end{subequations}
This pair of  coupled ordinary differential equations provides an arc-length-parametetrized solution of the EL equation. 
In this Cartesian presentation,  it is evident that translations amount to redefinitions of the parameters.
\\\\
An alternative derivation of Eq.(\ref{GYXprime}a) is instructive.
This involves cycling the integration of the conserved tension, captured by 
Eqs.(\ref{FX}) and (\ref{FY})  through the EL equation expressed in terms of the conformal curvature, Eq.(\ref{EperpK}):
$ -\mu\mu''- \kappa^2\mu^2/2  +  (\mu')^2/2= \mathcal{K}_0$, 
where $\mathcal{K}_0$ is the constant conformal curvature.
Eliminating $\mu$ using Eqs.(\ref{FX}) and (\ref{FY})  
 Eq.(\ref{GYXprime})(a)  is recovered if the identification
$ \frac{1}{2} \mathcal{K}_0 +
G_1 F- M_1^2 + M_2^2 =0$ 
is made.  This is simply the equilibrium Casimir identity Eq.(\ref{Casimir}) in the appropriately rotated frame, with tension $F\,\mathbf{i}$, torque $M_2$ and scaling current $M_1$. 
If $F=0$,  so that Eq.(\ref{calCS}) can be used to express $\mathcal{K}_0$ as a function of $S$, the constraint on logarithmic spirals, $4M S= 1$, is reproduced.

\section{Generating Tension by inversions of log spirals}
\label{GeneratingTension}

Whereas the energies of two curves related by a conformal transformation coincide, the tension in the two will generally differ.  Because all equilibrium states are conformally equivalent to some tension-free logarithmic spiral, this provides a constructive approach to generating these states.  
\\\\
Begin by examining the behavior of  tension-free states under 
conformal inversion.  Using the results of section \ref{CTs}, the transformed tension (\ref{eq:Fkmu0}) reads
\begin{equation}
 \mathbf{F}\to \bar{\mathbf{F}} =|\mathbf{X}|^2 \mathrm{R}_\mathbf{X} \mathbf{F} 
+  |\mathbf{X}|^2\left(\frac{\mathbf{n}\cdot \mathbf{X}}
{|\mathbf{X}|^2 \kappa'{}^{1/2}}\right)'\, \mathrm{R}_\mathbf{X} \mathbf{t} 
+ 
 |\mathbf{X}|^2\left(\frac{\mathbf{t}\cdot \mathbf{X}}{|\mathbf{X}|^2 \kappa'{}^{1/2}}\right)'\, (-)\mathrm{R}_\mathbf{X}\mathbf{n}\,.
\end{equation}
In the first term, the shorthand $ \mathrm{R}_\mathbf{X} \mathbf{F} $ has been used,  where
$\mathrm{R}_\mathbf{X}$ is the linear operator defined by Eq.(\ref{eq:Rdef}) representing a reflection orthogonal to $\mathbf{X}$. 
Technically, a relative minus sign appears  in the normal projection associated with the orientation change of $\mathbf{n}$. 
\\\\
Let $\bar{\mathbf{F}}=0$, so that
\begin{equation}
\mathbf{F} =
- \left(\frac{\mathbf{n}\cdot \mathbf{X}}
{|\mathbf{X}|^2 \kappa'{}^{1/2}}\right)'\, \mathbf{t} 
+ 
\left(\frac{\mathbf{t}\cdot \mathbf{X}}{|\mathbf{X}|^2 \kappa'{}^{1/2}}\right)'\, \mathbf{n}\,.
\end{equation}
Projection onto tangential and normal directions, as in section \ref{IntegrateF}, gives 
\begin{equation}
\label{FxpYp}
FX' = - \left(\frac{\mathbf{n}\cdot \mathbf{X}}
{|\mathbf{X}|^2 \kappa'{}^{1/2}}\right)' \,,\quad 
FY'=\pm \left(\frac{\mathbf{t}\cdot \mathbf{X}}{|\mathbf{X}|^2 \kappa'{}^{1/2}}\right)' \,;
\end{equation}
which can be integrated. 
Squaring and summing the integrated expressions gives
\begin{equation}
|\mathbf{X}|^4\, |F\mathbf{X}+\mathbf{M}_0|^2  \,\kappa' =  (\mathbf{n}\cdot \mathbf{X})^2 + 
(\mathbf{t}\cdot \mathbf{X})^2 \,,
\end{equation}
or
\begin{equation}
|\mathbf{X}|^2\,|F \mathbf{X}+\mathbf{M}_0|^2  \,\kappa'  =1\,,
\end{equation}
where again $\mathbf{M}_0=(M_1,M_2)$.
Thus $\kappa'$ depends only on the position.
Alternatively,  eliminating $\kappa'$ treating the component equations symmetrically, 
\begin{equation} 
(\mathbf{n}\cdot\mathbf{X}) \, (FY+M_2) = (\mathbf{t}\cdot\mathbf{X}) \, (FX+M_1) \,.
\end{equation}
The completeness of $\mathbf{t}$ and $\mathbf{n}$ then implies 
\begin{equation}
 |\mathbf{X}|^2=\left[ 1 + \left(\frac{FX+M_1}{FY+M_2}\right)^2\right]\,  (\mathbf{t}\cdot\mathbf{X}) ^2 
\,,\end{equation}
This determines $(\mathbf{t}\cdot\mathbf{X})$ completely in terms of $\mathbf{X}$.
If $F=0$, the only solution is  $\mathbf{X}\cdot\mathbf{t}= c |\mathbf{X}|$.
In general, the curve is traced by 
\begin{subequations}
\label{rhophi}
\begin{eqnarray}
 \left[ 1 + \left(\frac{F \rho\cos\phi +M_1}{F \rho\sin\phi +M_2}\right)^2\right]\,  {\rho'}^2 &=&1\,;\\  
{\rho'}^2 + \rho^2 \,{\phi'}^2 &=&1\,.
\end{eqnarray}
\end{subequations}
These equations are completely equivalent to Eqs.(\ref{GYXprime})  expressed in
polar variables. 
The apparent discrepancy between the two approaches 
is resolved by introducing an appropriate translation so that the origins coincide.  
\\\\
More straightforwardly, we simply note that all
equilibrium states can be generated by inversion of a logarithmic spiral in a sphere.
Let the sphere be centered at the  point $ \rho_0 x_0 \,\mathbf{i}$ with a radius $R$. The logarithmic spiral, centered on the origin with constant scaling rate $S_0$ is given by  $\rho/\rho_0=  \exp(\phi/4S_0^2)$. Under inversion, it maps to the curve\footnote{\sf $\phi$ is not the polar angle on the double spiral. $x_0$ here is adimensional, a ratio of scales.}
\begin{equation}
\label{Xdouble}
\rho_0 \mathbf{X}/R^2  = \frac{\left( \exp(\phi/4S_0^2) \,\cos \phi - x_0,   \exp(\phi/4S_0^2)\, \sin \phi\right)}{\left( \exp(\phi/4S_0^2)\, \cos \phi - x_0\right)^2 + \left(\exp(\phi/4S_0^2)\, \sin \phi\right)^2 }\,.
\end{equation}
To avoid dimensionally correct clutter, we will set $R=1$ and $\rho_0=1$. 
If $\phi\to -\infty$, $\mathbf{X}\to (1,0)/x_0$; whereas if $\phi\to \infty$, $\mathbf{X}\to \mathbf{0}$. Thus the distance between the two poles, centered along the $x$-axis is $x_0^{-1}$. Each of these geometries is a double spiral.  Notice that the chirality changes upon inversion.  
\\\\
If $x_0=- (-1)^n \exp(n\pi / 4S_0^2)$, with $n$ an integer,
the geometry is a symmetric double spiral with $Z_2$ rotational symmetry about the point
$(x_0/2,0)$.  This geometry is illustrated when $x_0=-1$ ($n=0$) as the black curve in Figure 3.
\\\\
If $x_0= (-1)^n \exp(n\pi / 4S_0^2)$,  
the center of inversion lies on the logarithmic spiral. Points on the two sides get mapped to infinity in opposite directions.
The double spiral disconnects into two parts. The asymptotic line of this hyperbolic spiral has a tangent angle $\theta_\infty$, 
independent of $x_0$ and  the order of intersection $n$, given by 
$\tan \theta_\infty= 2S_0^2$. This is easily confirmed: notice that, when $e^{2\pi v n}= x_0$, the numerators in Eq.(\ref{Xdouble})
vanish so the the tangent vector is determined by their derivatives. 
Again these limiting geometries exhibit $Z_2$ symmetry.   
The conserved currents are continuous as the parameters are dialed through the hyperbolic values. 
\\\\
For fixed $S_0$,  dialing $x_0$, we alternate between symmetric double and hyperbolic spirals.
Notice that while the value $\phi=n\pi $ occurs along the horizontal axis, $\phi=0$ itself has no geometrical significance. For each geometry $\kappa$ 
is monotonic, exhibiting a single inflection point (or linear asymptotes, if hyperbolic) with $\kappa=0$. Generally this inflection point does not occur on the horizontal axis, except when $x_0$ is tuned so that the geometry possesses $Z_2$ symmetry and it occurs at $(x_0/2,0)$. 
\\\\
The constant tension, given by Eq.(\ref{eq:Fkmu0}), is evaluated for the double spiral to be 
$\mathbf{F}=  x_0 \, (4 S_0^2 \, \mathbf{i} -  \mathbf{j} )/ (2S_0)$.
The magnitude is proportional to $x_0$ (the inverse distance between the poles), with a minimum, for fixed $x_0$  when $2S_0=1$, so that the conformal curvature, given by  Eq.(\ref{calCS}), vanishes. Interestingly, vanishing conformal curvature does not imply vanishing $\mathbf{F}$. $|\mathbf{F}|$ diverges at $S_0=0$ and well as in the limit, $S_0\to \infty$.
\begin{figure}[htb]\begin{center}
\subfigure[]{\includegraphics[height=3cm]{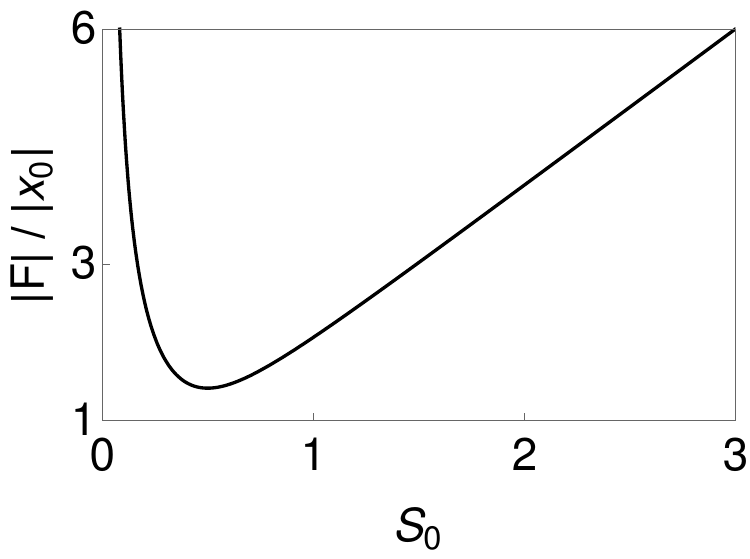}}\hskip.5cm
\subfigure[]{\includegraphics[height=3cm]{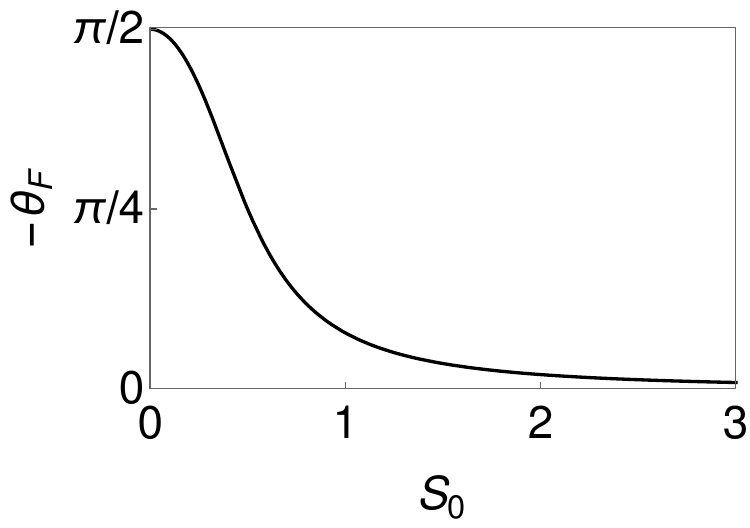}}
\caption{\small  Behavior of $|\mathbf{F}|$ and $\mathbf{F}\cdot \mathbf{i}$ as functions of $S_0$ for fixed values  of $x_0$.}
\label{Fig:FSX0}
\end{center}
\end{figure}
Notice that the tension is never directed along the line connecting the two poles. 
\begin{figure}[htb]
\begin{center}
\subfigure[]{
\includegraphics[height=3cm]{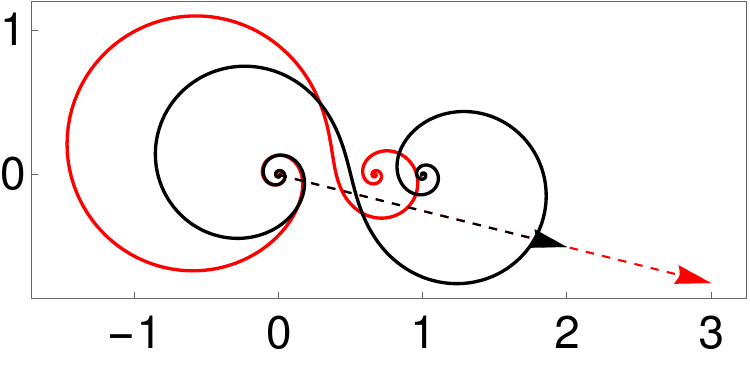}}
\hskip.5cm
\subfigure[]{\includegraphics[height=3.5cm]{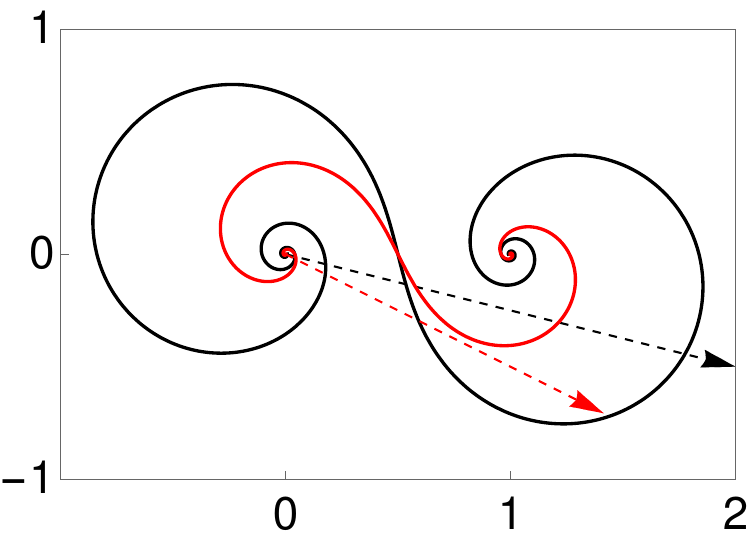}}
\caption{\small \sf Double Spiral trajectories $\mathbf{F}$ for fixed $S_0$ and fixed $x_0$:
(a) $S_0=1$, Black $x_0=-1$, Red $x_0=-3/2$; (b) $x_0=-1$, Black $S_0= 1$, Red $S_0=1/\sqrt{2} $}\label{Fig:TFSX0}
\end{center}
\end{figure}
The tangent of the angle $\mathbf{F}$ makes with this direction is $\tan \theta_\mathbf{F}=-1/4S_0^2$, independent of $x_0$, and as such completely determined by the  original logarithmic spiral. It is not a coincidence that $\theta_\mathbf{F}$ coincides with the angle that the normal to this spiral makes with the radial direction discussed below Eq.(\ref{Xs}): this is, after all, the only 
angle defined for any fixed $S_0$.  
In this context it is also relevant that the 
chiral double spirals are not symmetric with respect to reflections in the $Y$ axis. The offset $\mathbf{F}$ with respect to the vector connecting the poles reflects this asymmetry. Its value, of course, is encoded in the scaling behavior of the original logarithmic spiral.    
This vector is illustrated in Figures \ref{Fig:TFSX0} (a) and (b) for fixed $S_0$ and fixed  $x_0$ respectively.

\section{Tension-free states of higher-order invariant energies}
 \label{linKappa}

It is instructive to examine how the framework extends to higher-order conformally invariant energies.
The simplest extension is one linear in $\mathcal{K}$,
\begin{equation}
 H=\int\,ds\,\mu^{-1}\left(\mathcal{K}+\beta\right)\,.
\end{equation}
One can show that $S_D$, defined by Eq.(\ref{eqLSDdefN2}), is given by
\begin{equation}
S_D = -\frac{\kappa\mu}{2}\left(\mathcal{K}-\beta\right)-\mu'\,.
\label{SdK}
\end{equation}
As a consequence, Eq.(\ref{ELSD}) implies
that 
\begin{equation}\label{ELSDK}
 \mu^2\mathcal{E}_\perp= \frac{3}{2}\mathcal{K}_1\mathcal{K}-\frac{1}{2}\mathcal{K}_3 - \beta \frac{1}{2} \mathcal{K}_1=\mathcal{J}_1\,,
\end{equation}
where the subscript $N$ has been introduced to denotes the $n^{\sf th}$ derivative with respect to conformal arc-length, and $\mathcal{J}$ is given in the manifestly conformally invariant form, 
\begin{equation}
\label{Jdef}
 \mathcal{J}=-\frac{1}{2}\left(\mathcal{K}_2-\frac{3}{2}\mathcal{K}^2\right) -  \beta \frac{1}{2} \mathcal{K}\,.
\end{equation}
$\mathcal{J}$ is the Casimir invariant of the conformal group for this this linear energy.
The details of the calculation will be presented elsewhere in a more general context. 
In equilibrium $\mathcal{J}$ is a constant.  This can be integrated to express $\mathcal{K}$ in terms of Elliptic functions. As before, the focus will be on tension-free states so this is not the route we will take. 
\\\\
In tension-free states, the scaling current and torque are given respectively by
\begin{subequations}
\label{SM}
\begin{eqnarray}
 S &=& -\frac{\kappa\mu}{2}\left(\mathcal{K}-\beta\right)-\mu'\,;\\
 M&=&-\frac{\mu'}{2}\left(\mathcal{K}-\beta\right)-\frac{1}{2}\mathcal{K}_1+\kappa\mu.
\end{eqnarray}
\end{subequations}
As before, logarithmic spirals exist with $\kappa= 4s_0^2/s$, and constant 
$\mathcal{K}= (1- 16s_0^4)/(8s_0^2)$;  however, 
$s_0$ is no longer the scaling current. 
The values of $S$ and $M$ are given, for these logarithmic spirals, in the parametric form 
\begin{equation}
 S=\left(2s_0^3+\frac{3}{8s_0}\right)+\beta s_0\,;
 M=\left(\frac{3s_0}{2}+\frac{1}{32s_0^3}\right)-\frac{\beta}{4s_0}\,,
\end{equation}
representing the curve on the $(S,M)$ plane exhibiting a cusp when $s_0=1/2$
 illustrated in Figure \ref{phi0} (with $\beta=0$).
\begin{figure}[htb]
\begin{center}
{\includegraphics[width=5cm,height=5cm]{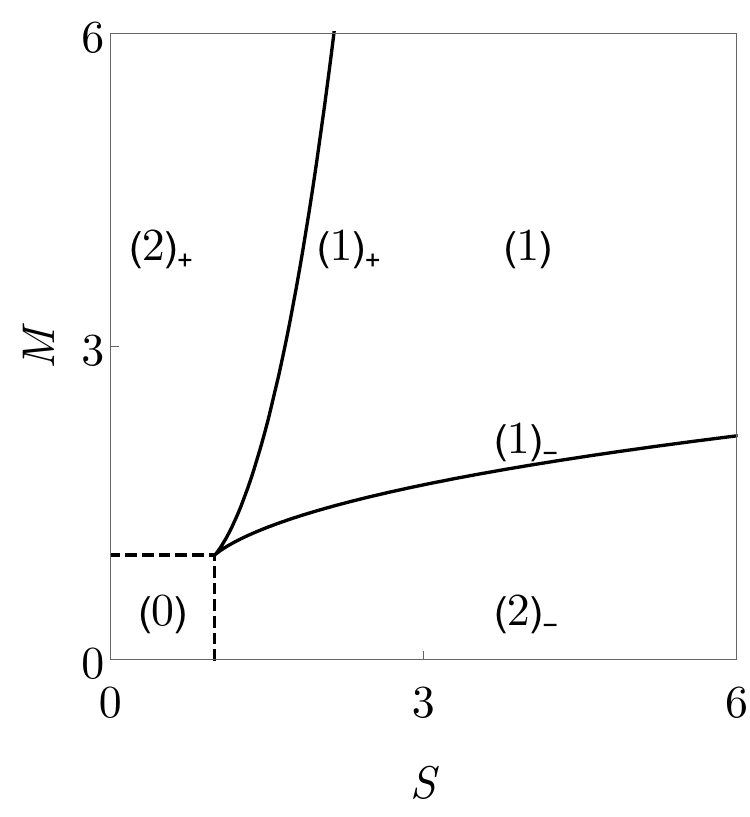}}
\end{center}
\caption{
The values of $S$ and $M$ representing  logarithmic spirals occur along the black curve.
If  $\beta=0$, this curve  is symmetrical with respect to $S$ and $M$.  In general, however it is not.  The upper (lower) branch (labelled $(1)_+$ ($(1)_-$)), corresponds  
to spirals with $s_0<1/2$ so that $\mathcal{K}>0$ ($s_0<1/2$ and $\mathcal{K}<0$); the two meet at a cusp  at the point $(1,1)$ ($s_0=1/2$) representing spirals with 
$\mathcal{K}=0$.  The Frenet curvature $\kappa=4s_0^2/s$ increases as one moves from left to right.
The quadrant is partitioned into four regions denoted $(0)$, $(1)$, $(2)_-$ and $(2)_+$ by this curve, the  significance of each region is described in the text.}
\label{MvsS}
\end{figure}
In general, the scale invariance captured by Eq.(\ref{SM}a) implies a non-linear ode third-order in derivatives of $\kappa$; 
rotational invariance (\ref{SM}b) implies an ode, one order higher still. These are two orders higher than their 
counterparts for conformal arc-length. Clearly, logarithmic spirals are not the  only tension-free states. Each point in $(S,M)$ space represents a unique tension-free state, a very different state of affairs compared to conformal arc-length.

\subsection{A quadrature}

\begin{figure}[htb]
\begin{center}
\subfigure[]{\includegraphics[width=2.5in,height=5cm]{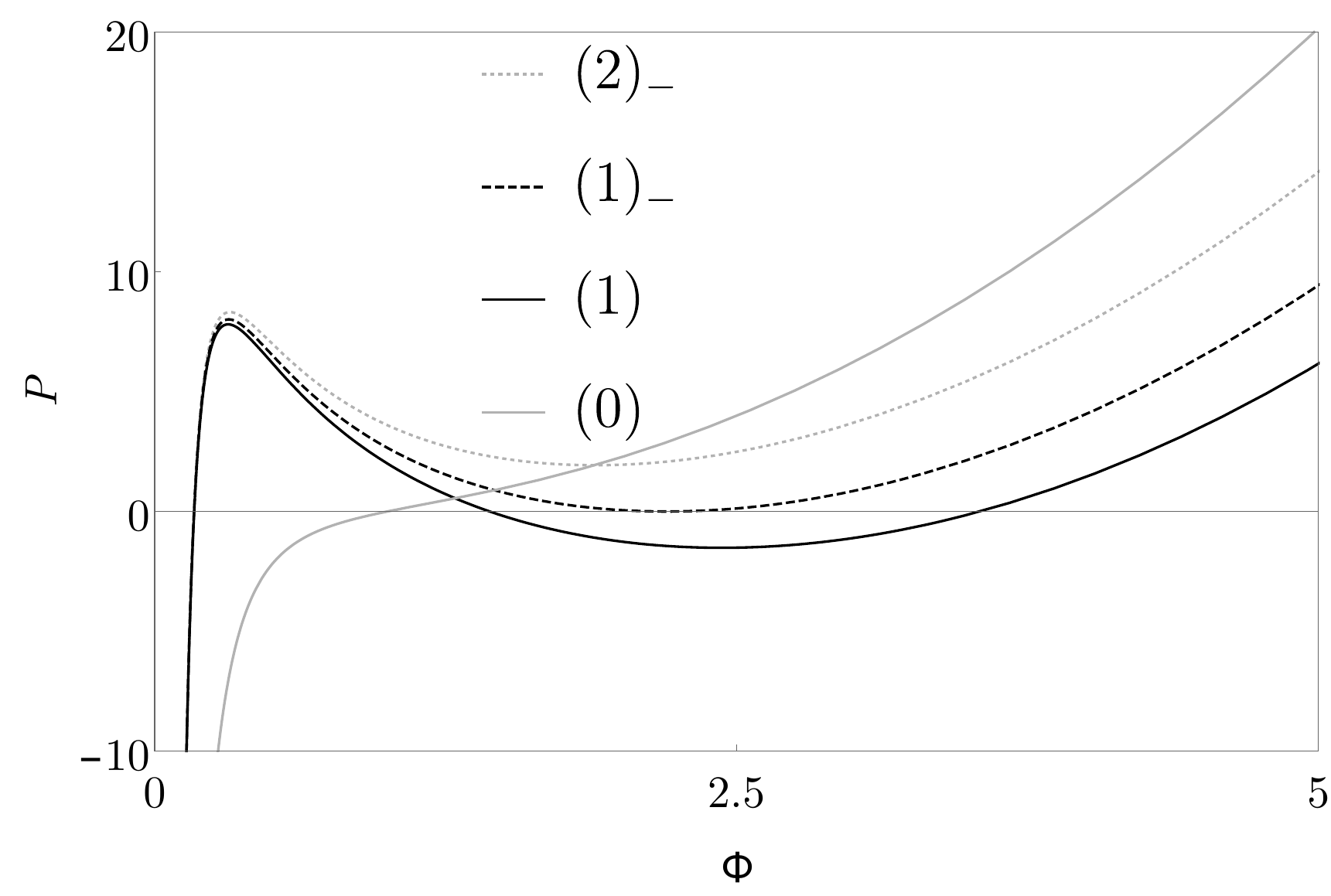}\label{PotencialA}}
\hspace{1cm}
\subfigure[]
{\includegraphics[width=2.5in,height=5cm]{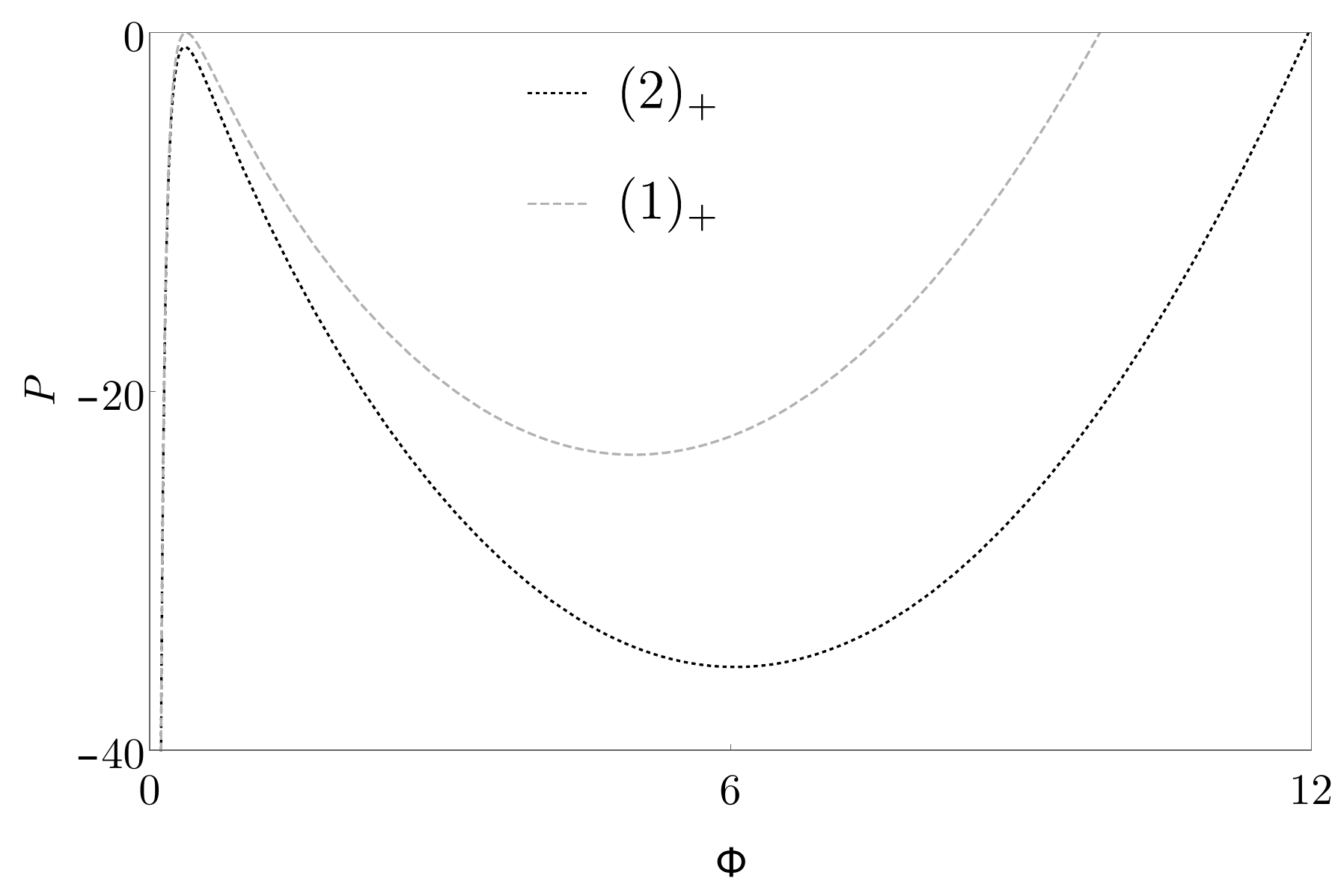}\label{PotencialB}}
\caption{Potential $P$ for couples $(S,M)$ in different regions of parameter space (cf. Figure \ref{MvsS}). In Figure \ref{PotencialA}, $S=3$.  If $M=6/5$ the well is inaccessible (in region $(2)_-$); if $M=1.67$, the  minimum occurs where $P=0$ representing a logarithmic spiral ($(1)_-$) ; if $M=2$ $\Phi$ oscillates between two turning points (region $(1)$).  If $(S,M)=(1/2,1/2)$, the well is monotonically increasing from zero with no well (region $(0)$). In Figure \ref{PotencialB},  $S=2$: $M=4.92$ corresponds to a point lying along 
$(1)_+$ (admitting a logarithmic spiral); if $M=6$ the well is accessible but the lower turning point lies at $\Phi=0$ (region 
$(2)_+$). }
\label{Potential}
\end{center}
\end{figure}

One way to integrate Eqs.(\ref{SM}) is the following:  use $S'=0$ in Eq.(\ref{SM}a) to eliminate $\mathcal{K}'$ in Eq.(\ref{SM}b). The latter can now be cast 
\begin{equation}\label{SMb1}
 \kappa\mu M-\mu\mu''=\frac{1}{2}\left(\mathcal{K}-\beta\right)+\kappa^2\mu^2\,.
\end{equation}
It is useful to introduce the dimensionless variable 
$\Phi= \kappa\mu$.\footnote{\sf
Note that $\kappa^2+\kappa'\Phi^2=0$, so that $\kappa$ is monotonic in arc-length.} 
Now $\mu'= \dot \Phi - 1/\Phi$, where dot represents a derivative with respect to the rotation angle, $\kappa^{-1} d/ds$, and
\begin{equation}
\label{KappaPhidots}
 \mathcal{K}=-\Phi\ddot\Phi-\frac{2\dot\Phi}{\Phi}-\frac{\Phi^2}{2}+\frac{\dot\Phi^2}{2}+\frac{1}{2\Phi^2}\,.
\end{equation}
It is now possible to recast Eqs.(\ref{SM}a) and (\ref{SMb1})  in terms of the variable $\Phi$ and its first two derivatives with respect to the rotation angle:
\begin{equation}\label{SMa1}
\frac{1}{2}\Phi\ddot{\Phi}-\frac{\dot{\Phi}^2}{4}+\frac{\Phi^2}{4}+\frac{3}{4\Phi^2}+\frac{\beta}{2}-\frac{S}{\Phi}=0\,;
\end{equation}
and
\begin{equation}\label{SMb2}
\frac{1}{2}\Phi\ddot{\Phi}+\frac{\dot{\Phi}^2}{4}+\frac{3\Phi^2}{4}+\frac{1}{4\Phi^2}-\frac{\beta}{2}-M\Phi=0
\end{equation}
respectively. 
Taking the difference (\ref{SMb2})-(\ref{SMa1}) to eliminate $\ddot{\Phi}$, a  simple quadrature 
\begin{equation}\label{Quad1}
\dot{\Phi}^2+\Phi^2-\frac{1}{\Phi^2}+2\left(\frac{S}{\Phi}-M\Phi-\beta\right)=0
\end{equation}
for $\Phi$ is identified, which can be interpreted in terms of a particle in a potential.
The second linear combination,  (\ref{SMb2})+ (\ref{SMa1}), yields
\begin{equation}\label{Phi2dot1}
 \Phi\ddot{\Phi} +\Phi^2+\frac{1}{\Phi^2}- \left( \frac{S}{\Phi} + M\Phi \right) =0\,,
\end{equation}
reproducing the quadrature upon integration. Consistency requires the identification of the constant of integration with $\beta$.
\\\\
Focus now on the case $\beta=0$.
The potential $P=\Phi^2-\frac{1}{\Phi^2}+2\left(\frac{S}{\Phi}-M\Phi\right)$, 
appearing in the quadrature is plotted for representative values of $S$ and $M$ in Figure \ref{Potential}.
In region $(1)$ of $(S,M)$ space, there is always an admissible potential well, of finite depth, with turning points at finite (non-vanishing) values of $\Phi$.
The potentials correspond to the boundaries of this region, $(1)_-$ and $(1)_+$ are illustrated in Figure \ref{Potential}: in the former case, $P=0$ at the bottom of the well; in the latter $\dot \Phi=0$ at the left hand turning point. In both cases, $\Phi$ is constant, representing logarithmic spirals.  Outside region  $(1)$ there is no finite well. In  $(1)$ there is a second monotonic infinite well extending from $\Phi=0$ to the first zero of the potential coexisting with the bounded well; in $(0)$, $(2)_-$ and $(2)_+$ the only accessible well extends from $\Phi=0$, where it diverges as $-\Phi^{-2}$ independent of the parameter values, to the single zero of $P$.  In $(2)_+$, as is evident in Figure \ref{Potential}, this well is not monotonic, in contrast to its monotonic form elsewhere in parameter space.

\begin{figure}[htb]
\begin{center}
\includegraphics[width=2.5in,height=5cm]{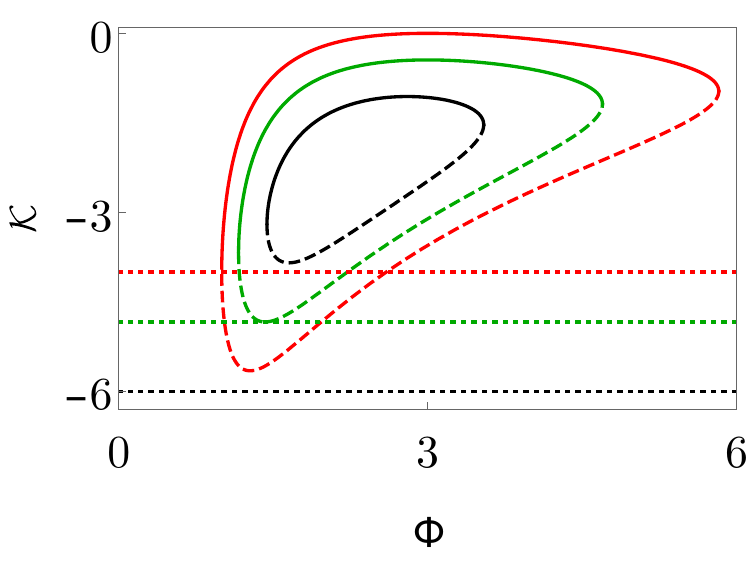}
\end{center}
\caption{Three  $\mathcal{K}$ cycles  for  fixed $S=3$, and $M=2$ (black), $M=2.48$ (green) and $M=3$ (red). The solid curves corresponds to ${\sf sign}({\dot\Phi})>0$;  the dashed line to negative values. The dotted horizontal lines represents $\mathcal{K}=-4S/M$ for 
the respective values of $S$ and $M$. The intersections of each line with the corresponding cycle indicate the values of $\Phi$ where the distance to the spiral apex exhibits a local maximum or minimum ($\rho'=0$), 
as described in section \ref{CCurrent}.}
\label{KvsPhi(1)}
\end{figure}

\noindent It is easy to see that the radius of curvature $\kappa^{-1}$ has an inflection point 
at the turning points of the potential. 
\\\\
Of special interest are the oscillatory trajectories realized in region $(1)$ (cf. Figure \ref{Potential}). 
Using Eq.(\ref{KappaPhidots}) and the quadrature (\ref{Quad1}), it is possible to express the 
conformal curvature in terms of $\Phi$ and ${\sf sign}(\dot\Phi)$:
\begin{equation}\label{EqKPhi}
 \mathcal{K}=\frac{2}{\Phi^2}-\frac{2S}{\Phi} - {\sf sign} (\dot\Phi) \frac{2\sqrt{-P(\Phi)}}{\Phi}\,.
\end{equation}
For oscillatory trajectories, this equation describes bounded closed cycles on the $(\Phi,\mathcal{K})$ plane as illustrated in Figure \ref{KvsPhi(1)};
the upper (lower) half cycle corresponding to $\dot\Phi<0$ ( $>0$).   
$\mathcal{K}$ will change sign twice along the cycle whenever $\Phi^2-2M\Phi+S^2=0$. This is only possible if $M\geq S$. 

\subsection{The special conformal current and the distance to the spiral apex} 
\label{CCurrent}

\begin{figure}[htb]
\begin{center}
\includegraphics[width=2.5in,height=5cm]{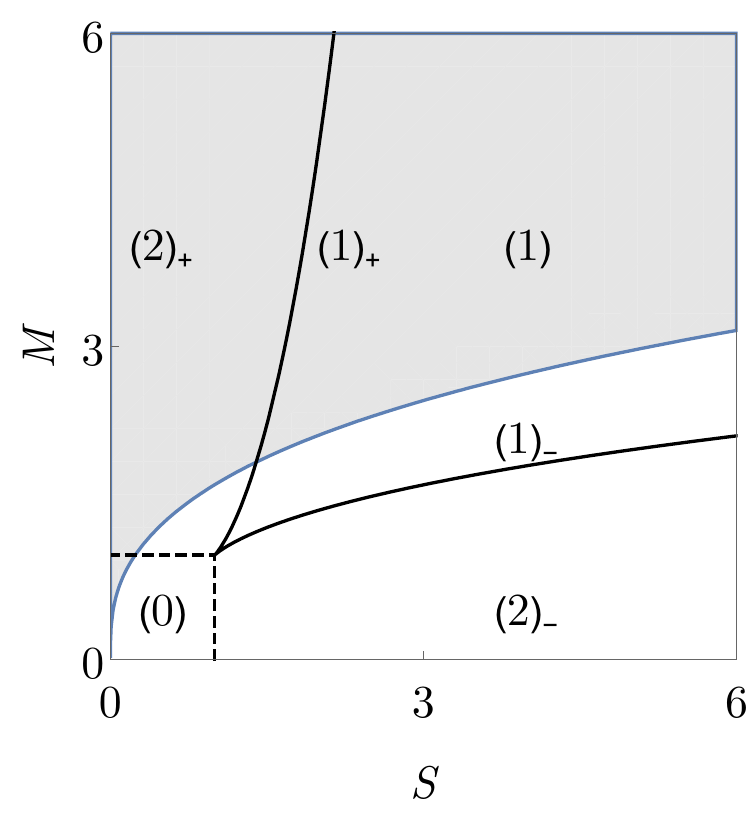}
\end{center}
\caption{$\rho'>0$ is everywhere positive in the white region; It is negative in places within the gray region.  Region $(1)$ is thus partitioned into monotonic and non-monotonic regions. The former 
includes $(1)_-$.  One can check that, within region $(1)$,  all values of $\Phi$ where $\rho'<0$ 
lies within the in the potential well.}
\label{RhoPrime0}
\end{figure}

The special conformal current (\ref{eq:Gdef}) is given for the theory linear in $\mathcal{K}$ 
by the expression:
\begin{equation}
\label{EcG0}
 \mathbf{G}=2M\mathbf{X}_\perp-|\mathbf{X}|^2\mathbf{F}+2S\mathbf{X}+\mu\mathcal{K}\mathbf{n}+4\mu\,\mathbf{t}\,.
\end{equation}
With respect to the spiral apex, $\mathbf{G}=0$ in any tension-free state. 
Its projection onto the orthogonal vectors, $\mathbf{X}$ and $\mathbf{X}_\perp$, then give respectively
\begin{equation}\label{EqSsqare}
 -2S|\mathbf{X}|^2=\mu\mathcal{K}(\mathbf{n}\cdot\mathbf{X})+4\mu(\mathbf{t}\cdot\mathbf{X})\,;
\end{equation}
\begin{equation}\label{EqMsqare}
 -2M|\mathbf{X}|^2=\mu\mathcal{K}(\mathbf{t}\cdot\mathbf{X})-4\mu(\mathbf{n}\cdot\mathbf{X})\,.
\end{equation}  
Taking linear combinations to eliminate the projections of $\mathbf{X}$ on $\mathbf{t}$ and $\mathbf{n}$ respectively, implies  
\begin{equation}\label{EcDPhi}
 (\mathcal{K}S-4M)|\mathbf{X}|^2+\frac{\mu}{2}(\mathcal{K}^2+16)(\mathbf{n}\cdot\mathbf{X})=0\,,
\end{equation}
and 
\begin{equation}\label{EcDRho}
 (\mathcal{K}M+4S)|\mathbf{X}|^2+\frac{\mu}{2}(\mathcal{K}^2+16)(\mathbf{t}\cdot\mathbf{X})=0\,.
\end{equation}
The quotient of these two equations implies 
\begin{equation}
 \rho'^2=\frac{1}{1+\left(\frac{\mathcal{K}S-4M}{\mathcal{K}M+4S}\right)^2}\leq1\,.
\end{equation}
Using the completeness of $\mathbf{t}$ and $\mathbf{n}$,  on the other hand, they imply the simple local relationship between $\rho$ and $\mathcal{K}$:
\begin{equation}\label{EqRho2}
 4(S^2+M^2)\rho^2=\mu^2\left(\mathcal{K}^2+16\right)\,.
\end{equation}
When $\mathcal{K}M+4S=0$,  $\rho'=0$, so that the polar radius $\rho$ is not monotonic.  Using Eq. (\ref{EqKPhi}),  it is evident that this value of $\mathcal{K}$ is realized 
when $\Phi$ satisfies
\begin{equation}\label{Poly2Rho0}
 \left(\frac{4S^2}{M^2}+1\right)\Phi^2-\left(\frac{4S^2}{M}+2M\right)\Phi+\left(S^2+\frac{4S}{M}\right)=0\,,
\end{equation}
with real solutions if and only if 
\begin{equation}
 \left(\frac{4S^2}{M}+2M\right)^2\geq4\left(\frac{4S^2}{M^2}+1\right)\left(S^2+\frac{4S}{M}\right)\,.
\end{equation}
This domain is illustrated in gray in Figure \ref{RhoPrime0}.

\noindent Note that $\rho'=1$ if and only if $\mathcal{K}S-4M=0$. One can show that in region $(1)$, the corresponding values of $\Phi$ are inaccessible.

\subsection{Constructing decorated logarithmic spirals}

Let us first examine small deformations of logarithmic spirals
\\\\
\textbf{$(1)_-$:}  In the neighborhood of region $(1)_-$ lying within region $(1)$ there exist tension-free states represented by small oscillations about the minimum of the potential.  One can now use the harmonic approximation to determine the evolution of $\Phi$: for given $s_0$,
\begin{equation}
 \ddot\Phi+\frac{3}{2}\left(1-\frac{1}{16s_0^4}\right)(\Phi-2s_0)\approx0\,,
\end{equation}
with period in the rotation angle given by $\Theta_0=2\pi/\omega$, where $\omega=\sqrt{3/2}\sqrt{1-1/(2s_0)^4}$. Recalling that $ \phi=\Theta$ in a logarithmic spiral (cf. footnote \ref{Thetaphi}), one finds that 
in each period $\Theta_0$, $\phi$ rotates by the angle $\phi_0=8\pi s_0^2\sqrt{2/3}/\sqrt{16s_0^4-1}$,
diverging both at the cusp point ($s_0=1/2$) as well as at large $s_0$.
\\\\
 \textbf{$(1)_+$:} There are no small oscillations about such logarithmic spirals, corresponding as they do to a maximum of the potential. 
\\\\
Now  let us examine the behavior in region $(1)$ more closely. 
\begin{figure}[htb]
\begin{center}
\subfigure[]
{{\includegraphics[width=2.0in,height=4cm]{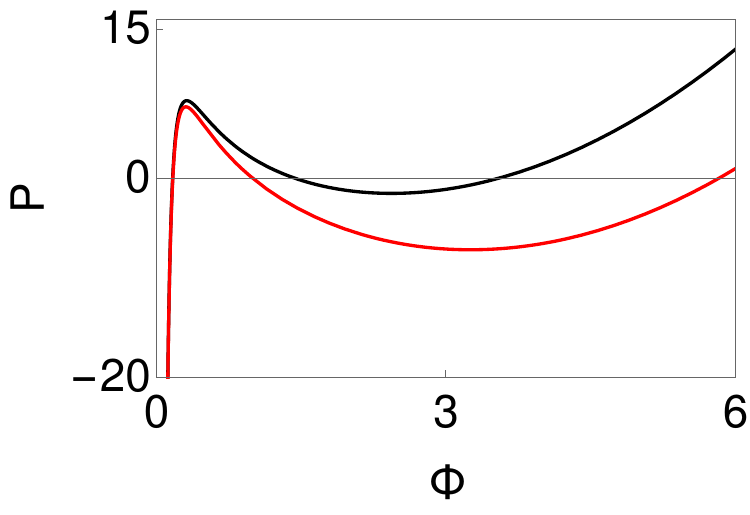}}}
\hspace{1cm}
\subfigure[]
{{\includegraphics[width=2.0in,height=4cm]{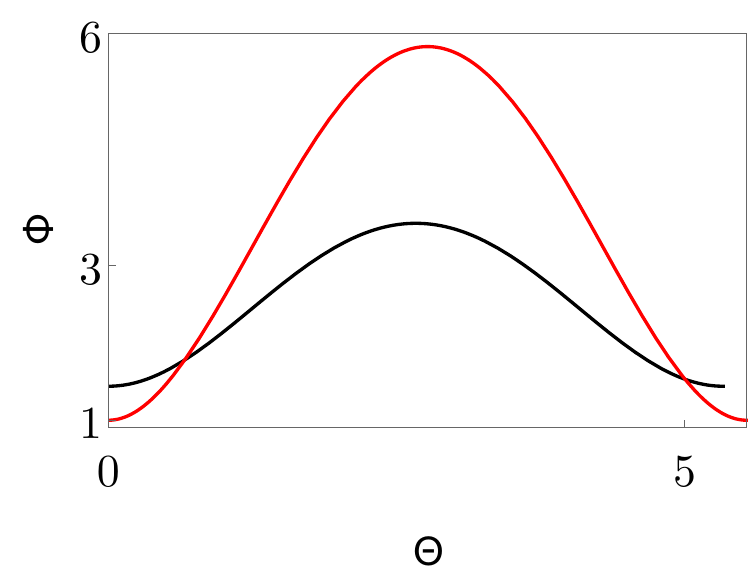}\label{PhiRegion1}}}
\caption{(a) Potential for $(S,M)=(3,2)$  (black) $\rho$ monotonic; $(S,M)=(3,3)$ (red) $\rho$ non-monotonic.  (b) Corresponding behavior of $\Phi$ over single period initialized at lower turning point.}
\end{center}
\end{figure}
\noindent 
One first solves the quadrature (\ref{Quad1})  within the well to determine $\Phi$ (cf. Figure \ref{PhiRegion1}). We do this for one set of parameter values representing a monotonically increasing spiral; and another representing one that is not.
The corresponding Frenet curvatures $\kappa$ (determined by the  differential equation $\kappa+\dot{\kappa}\Phi^2=0$) and conformal curvatures $\mathcal{K}$ (determined by Eq. (\ref{KappaPhidots}) are illustrated in Figure \ref{KappaGreen}) and  
Figure \ref{ConformalK} respectively. Notice that $\mathcal{K}$ is not symmetric with respect to
$\dot\Phi>0 \leftrightarrow \dot\Phi>0$ (cf. Eq.(\ref{EqKPhi}
)).
\begin{figure}[!htb]
\begin{center}
\subfigure[]
{\includegraphics[width=2.0in,height=4cm]{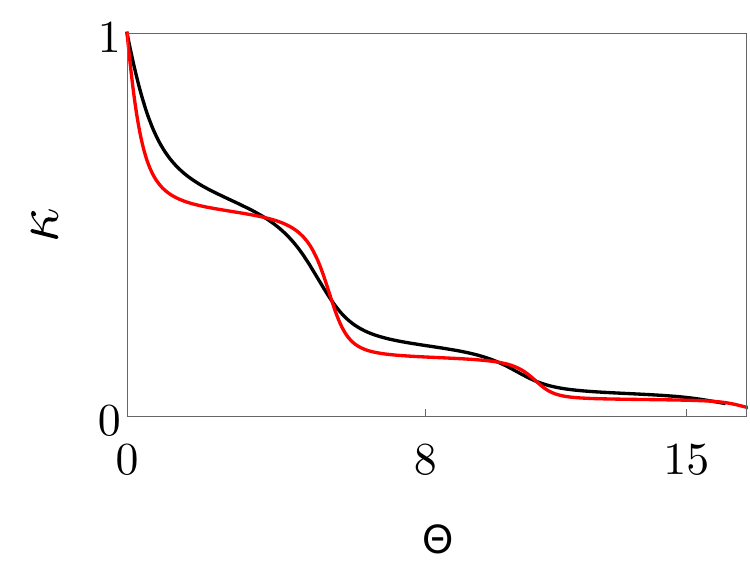}\label{KappaGreen}}
\hspace{1cm}
\subfigure[]
{\includegraphics[width=2.0in,height=4cm]{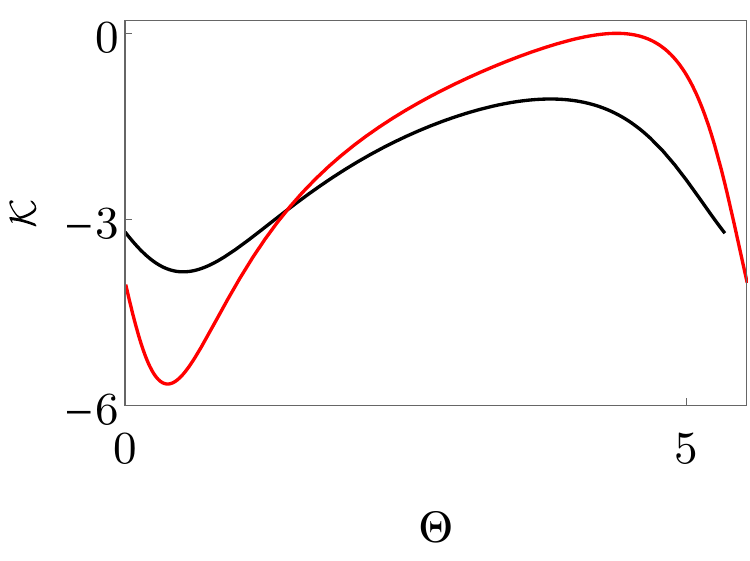}\label{ConformalK}}
\caption{(a) Frenet curvature for three periods of $\Theta$. Monotonic $(S,M)=(3,2)$ (black) and non-monotonic $(S,M)=(3,3)$ (red). The initial condition $\kappa(0)=1$ is used in both cases; (b) Single period of the conformal curvature for each case. Note the asymmetry reflecting ${\sf sign}\, \dot\Phi$. }
\end{center}
\end{figure}                                                                                                                     
\noindent 
The corresponding spirals can now be constructed from $\kappa$, or alternatively from Eq.(\ref{EqRho2}).
\begin{figure}[htb]
\begin{center}
\includegraphics[width=5cm,height=5cm]{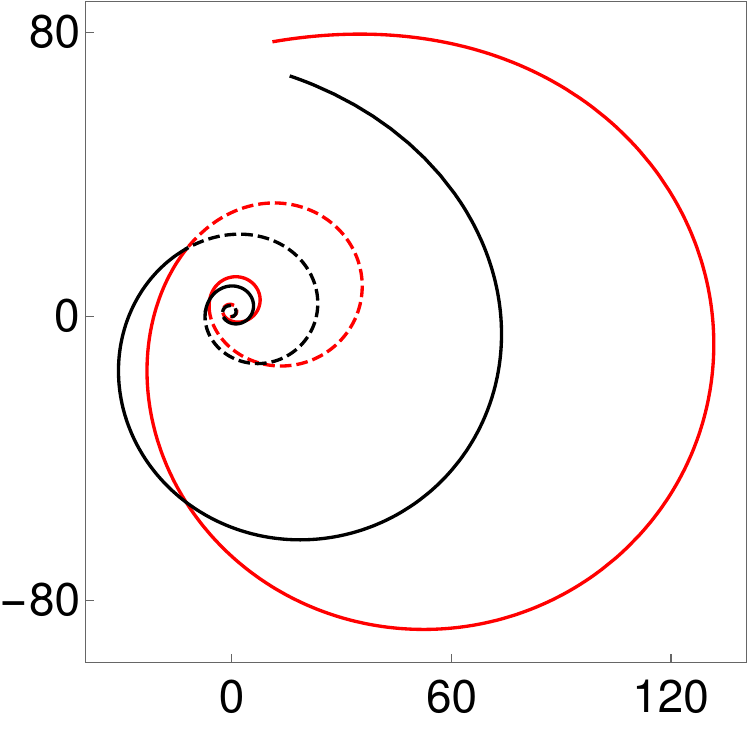}
\end{center}
\caption{Three periods of decorated logarithmic spirals for $S=3$: $M=2$ monotonic $\rho$ (black); $M=3$ non-monotonic $\rho$ (red). }
\label{phi0}
\end{figure}

\newpage
\subsection{$S=0$ and $M=0$}

In this case all conserved quantities vanish.  There are no infinite spirals, decorated or not, with these parameter values. What then does it represent?  
\begin{figure}[!htb]
\begin{center}
\subfigure[]{\includegraphics[width=4cm,height=4cm]{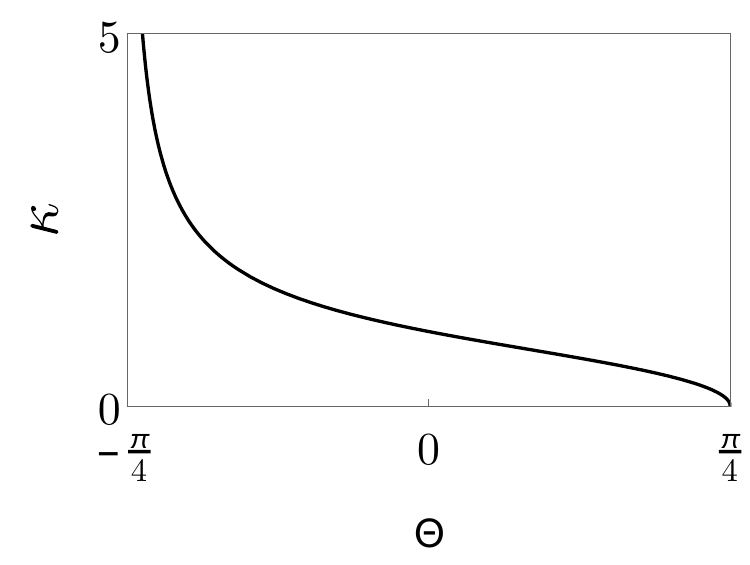}\label{CurvatureSyM0}}
\hspace{.3cm}
\subfigure[]
{\includegraphics[width=4cm,height=4cm]{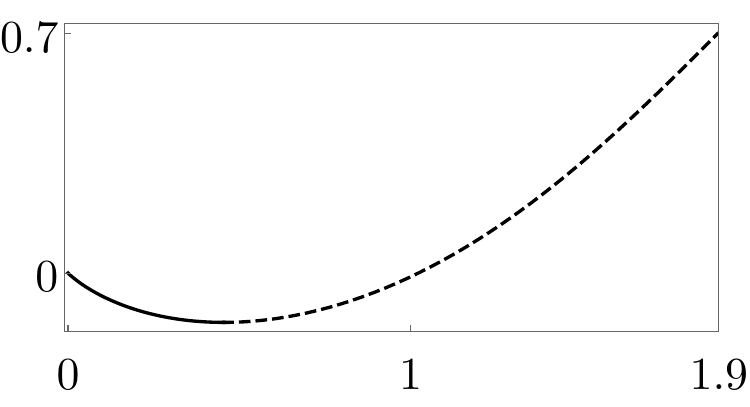}\label{CurveSyM0}}
\caption{(a) Curvature $S=M=0$; (b) Curve $S=M=0$. $\dot\Phi>0$ along solid black segment whereas 
$\dot\Phi<0$ along dashed segment.}
\end{center}
\end{figure}
\vskip1pc\noindent Now  $\dot{\Phi}^2=1/\Phi^2-\Phi^2$. If $\Phi$ is initialized at the turning point, $\Phi(0)=1$. The exact solution is then  $\arcsin(\Phi^2)=2\Theta+\pi/2$, or  
$\Phi^2=\cos2\Theta$, with $\Theta \in (-\pi/4,\pi/4)$. The equation $
\dot{\kappa}=-\kappa/\cos2\Theta$ is now solved to yield $\kappa=\kappa_0|\tan2\Theta+\sec2\Theta\,|^{-1/2}$ (cf. Figure \ref{CurvatureSyM0}).  $\kappa$ diverges at $\Theta=-\pi/4$ where $\Phi=0,\dot \Phi >0$ whereas 
$\dot \kappa$ diverges on the return of $\Phi$ to $\Phi=0$ with $\dot \Phi<0$ at $\Theta=\pi/2$.
The corresponding curve is rendered in Figure \ref{CurveSyM0}. Its length is finite. There is, however, no intrinsic length scale.  The notion that the \textit{ground state} is not a spiral is intriguing and something 
we worthy of further study. It should be borne in mind, however, that an analogue of this state exists for each point in parameter space. There are no non-trivial analogues of theses states in the minimal model.

\section{Conclusions}

In this paper we have shown how self-similar planar curves, and specifically logarithmic spirals, arise  as equilibrium states of conformally invariant energies. In this setting the characteristic self-similarity of Jacob Bernouilli's \textit{Spira Mirabilis}  can be attributed to its identification as a tension-free equilibrium state  of the simplest such energy.  As such, there is no intrinsic length scale. What is more, in this specific case, there is no internal structure: each point on a logarithmic spiral is equivalent modulo a similarity to any other point.
\\\\
The main focus has been on the simplest possible conformal mechanics of a planar curve  described by the conformal arc-length.  Any  other local conformally invariant energy necessarily involves higher-order 
derivatives of the curvature $\kappa$; the conformally invariant bending energy (\ref{eq:HB}), quadratic in the conformal curvature, involves three derivatives of $\kappa$.  Just as logarithmic spirals form the one-parameter family of tension-free states of the conformal arc-length,  each conformally invariant energy will exhibit its own self-similar tension-free states. 
Among them are logarithmic spirals whenever 
the scaling rate and the torque are tuned appropriately; however they do not exhaust the possibilities.
A brief sketch has been provided  for the simplest extension, linear in the conformal curvature, all the more simple because it is linear in $\kappa'''$.  It was shown how 
the mechanical point of view detailed for conformal arc-length is extended to construct the corresponding tension-free states.  Such states need not be infinite, but all infinite self-similar states are 
described by logarithmic spirals decorated by a repeating internal structure.  
Conformal arc-length is thus a limited guide to the general behavior. 
One would expect that additional levels of structure will be exhibited as the number of derivatives increases in the energy, such as the conformally invariant 
$H_{dB}= \int d\mathcal{S} \, \left( \frac{d \mathcal{K}}{d\mathcal{S}} \right)^2$
where (recall) $\mathcal{K}_a= d \mathcal{K}/ d\mathcal{S}$, and $\mathcal{S}$ is the conformal arc-length.
Understanding how this additional structure overlays the elementary spiral growth is a line of research likely to be worth pursuing.
\\\\
In general, the EL equation can be cast in a manifestly conformally invariant form. The Casimir invariant of the conformal group is no longer identified with the conformal curvature but some higher order polynomial in its derivatives. This subject will be addressed elsewhere. Just as the  equilibrium states of the conformal arc-length fall  into conformal equivalence classes, each containing a logarithmic spiral,  an analogous property holds in general with log spirals replaced by tension-free states. 
\\\\
The original motivation for this work was the identification of tension-free states of the generalization of the conformal arc-length to three or higher-dimensions, given by
\begin{equation}
H_0=  \int ds \, ({\kappa'}^2 + \kappa^2 \tau_1^2)^{1/4}\,,
\label{eq:SD}
\end{equation}
where $\tau_1$ is the first Frenet torsion \cite{Sharpe1994,Musso1994}.  This energy has been addressed by  Musso (in three dimensions)
and extended by Magliaro et al to higher dimensions \cite{Musso1994,Magliaro2011}.
To identify self-similar states it is necessary to 
reexamine the problem with a focus on the tension, and how its vanishing propagates through the remaining conservation laws \cite{Paper2,Paper3}. This focus involves dismantling the conformal invariance in the way this was done here.  These states exhibit considerably more structure than their planar counterparts.  The role of the scaling current is to constraint the torsion in terms of the curvature;  
torque conservation providing a quadrature for $\kappa'/\kappa^2$. The vanishing special conformal current then permits the construction of the spatial trajectories in a spherical coordinate system adapted to the spiral apex and the torque axis. 
\\\\ 
These tension-free spirals  form a two-parameter family of states labelled by their torque $M$ and their scaling current $S$.  In a planar logarithmic spiral it was seen that $4MS=1$. If 
the spiral is supercritical with $4MS>1$, it is found to exhibits a pattern of nutation between two oppositely oriented cones, twisting and untwisting within each cycle,  while this cycle precesses as the spiral expands.  Such  spirals intersect all planes; as such they are the natural spatial analogues of the logarithmic spiral.
\\\\
Time lapsed videos illustrating the trajectory described by the growing tendril of a climbing plant illustrate this expanding precessing pattern of nutating cycles as the tendril explores its environment in search of a point of attachment.  A moment's thought suggests that the plant is solving a three-dimensional analog  of  \textit{the  search for the shore} described in the introduction.  This \textit{coincidence} will be addressed elsewhere.

\vskip3pc \noindent{\Large \sf Acknowledgments}

\vskip2pc \noindent  
Partial support from CONACyT grant no. 180901  is acknowledged.


\begin{thebibliography}{99}
\bibitem{obs}
Two geometrical consequences of the identity (\ref{EperpK}) are the following: 
\begin{equation}
\int ds\, \mu  \,\mathcal{E}_\perp = -\frac{1}{2} \, \int ds\, \mathcal{K}'\,,
\end{equation}
and thus this new invariant is completely determined by the boundary conformal curvature. In addition,   
$\int ds\, \sqrt{\mathcal{E}_\perp}$ is also a conformal invariant. This may not command our immediate attention; intriguingly, however, the conformal curvature $\mathcal{K}$ plays the same role in this invariant as the one played by $\kappa$ in the conformal arc-length.
\end{thebibliography}
\end{document}